\documentclass[floatfix,twocolumn,showpacs,preprintnumbers,amsmath,amssymb,pra,superscriptaddress,longbibliography]{revtex4-1}
\usepackage{color}
\usepackage[usenames,dvipsnames,svgnames,table]{xcolor}
\usepackage[colorlinks=true,linkcolor=blue,urlcolor=blue,citecolor=blue]{hyperref}
\usepackage{url}
\usepackage{mathtools}
\usepackage{graphicx}
\usepackage{dcolumn}
\usepackage{array}
\usepackage{lipsum}
\usepackage{bm}
\usepackage{subfigure}
\usepackage{amssymb}
\usepackage{multirow}
\usepackage{tabularx}
\usepackage{amsmath}
\usepackage{braket}
\usepackage{csquotes}
\graphicspath{{plots/}}
 \usepackage{lipsum}
\usepackage{mathrsfs}
\usepackage{MnSymbol}
\newcommand{\beq}{\begin{equation}}
\newcommand{\eeq}{\end{equation}}
\newcommand{\bea}{\begin{eqnarray}}
\newcommand{\eea}{\end{eqnarray}}

\begin{document}
%\underline{Draft version:}
\title{\emph{Ab initio} path integral Monte Carlo study of the 2D uniform electron liquid at finite temperatures}
\author{Tobias Dornheim}
\email{t.dornheim@hzdr.de}
\affiliation{Institute of Radiation Physics, Helmholtz-Zentrum Dresden-Rossendorf (HZDR), D-01328 Dresden, Germany}
\affiliation{Center for Advanced Systems Understanding (CASUS) at Helmholtz-Zentrum Dresden-Rossendorf (HZDR), D-02826 G\"orlitz, Germany}

\author{Fotios Kalkavouras}
\affiliation{Electromagnetics and Plasma Physics, Royal Institute of Technology (KTH), Stockholm, SE-100 44, Sweden}

\author{Paul Hamann}
\affiliation{Institute of Radiation Physics, Helmholtz-Zentrum Dresden-Rossendorf (HZDR), D-01328 Dresden, Germany}
\affiliation{Institut f\"ur Physik, Universit\"at Rostock, D-18057 Rostock, Germany}

%\author{Sebastian Schwalbe}
%\affiliation{Institute of Radiation Physics, Helmholtz-Zentrum Dresden-Rossendorf (HZDR), D-01328 Dresden, Germany}

%\author{Pontus Svensson}
%\affiliation{Institute of Radiation Physics, Helmholtz-Zentrum Dresden-Rossendorf (HZDR), D-01328 Dresden, Germany}

\author{Zhandos A.~Moldabekov}
\affiliation{Institute of Radiation Physics, Helmholtz-Zentrum Dresden-Rossendorf (HZDR), D-01328 Dresden, Germany}

\author{Jan Vorberger}
\affiliation{Institute of Radiation Physics, Helmholtz-Zentrum Dresden-Rossendorf (HZDR), D-01328 Dresden, Germany}

\author{Panagiotis Tolias}
\affiliation{Electromagnetics and Plasma Physics, Royal Institute of Technology (KTH), Stockholm, SE-100 44, Sweden}

\begin{abstract}
\noindent We present extensive \emph{ab initio} path integral Monte Carlo (PIMC) simulations of the two-dimensional uniform electron gas (2DEG), covering a broad range of density parameters $r_s=0.1,\dots,50$ and temperatures $\Theta=k_\textnormal{B}T/E_\textnormal{Fermi}=0.5,\dots,16$. This allows us to analyze various structural, linear density response and spectral properties. We find clear evidence of a \emph{roton-type} feature in the dynamic structure factor at strong coupling and intermediate wavenumbers. We also benchmark novel dielectric theory implementations for the 2DEG~[Kalkavouras \emph{et al.}~arXiv:2601.14989] for structural and spectral properties across the liquid phase diagram. The PIMC results can be used to benchmark existing theories and approximations, and guide the development of new methodologies.
\end{abstract}
\maketitle

\section{Introduction}

The uniform electron gas (UEG)---often also referred to as \emph{jellium} or as quantum one-component plasma---arguably constitutes the most important model system in theoretical physics, quantum chemistry, and related research fields~\cite{quantum_theory,loos,review,Ott2018}. Originally introduced as a simple model for "simple" metals (Li, Na, K, Rb, and Cs)~\cite{mahan1990many,quantum_theory}, it has emerged as the archetypal model of interacting quantum electrons in a plethora of contexts. Highly accurate ground-state quantum Monte Carlo (QMC) simulations of the three-dimensional UEG~\cite{Ceperley_Alder_PRL_1980,moroni,moroni2,Ortiz_PRB_1994,Ortiz_PRL_1999,Spink_PRB_2013} have been used as input for readily available parametrizations of a variety of its properties, including the exchange--correlation (XC) energy~\cite{vwn,Perdew_Zunger_PRB_1981,Perdew_Wang_PRB_1992}, local field correction~\cite{cdop}, and pair correlation function~\cite{Gori-Giorgi_PRB_2000,Gori-Giorgi_PRB_2002}. These achievements have been useful for a plethora of applications, most notably the remarkable success of density functional theory~\cite{Jones_RMP_2015}.

More recently, cutting-edge UEG research has focused on the 3D UEG at finite temperatures~\cite{review}. This has been motivated by the surge of interest in \emph{warm dense matter}, an extreme state that naturally occurs in astrophysical objects such as giant planet interiors, brown dwarfs and white dwarf atmospheres, and the outer layer of neutron stars, and which is also relevant for material science and material discovery experiments as well as for inertial confinement fusion; see Refs.~\cite{review,vorberger2025roadmapwarmdensematter,bonitz2024principles} for recent review articles. The pressing need for an accurate description of the warm dense UEG has led to a number of significant advances in thermal QMC simulation methods, most notably different flavors of the \emph{ab initio} path integral Monte Carlo (PIMC)~\cite{cep} method~\cite{Dornheim_NJP_2015,dornheim_prl,Schoof_PRL_2015,Xiong_JCP_2022,Dornheim_JCP_xi_2023,Yilmaz_JCP_2020,Dornheim_PRR_2026}, but also density matrix QMC~\cite{Blunt_PRB_2014,Malone_JCP_2015,Malone_PRL_2016}, auxiliary field QMC~\cite{Joonho_JCP_2021} and diagrammatic QMC~\cite{Hou_PRB_2022,Li_PRB_2025,hou2026firstprincipleseffectivemassthreedimensional}.
As a consequence, accurate results have become available for a broad class of observables, such as different energies, free energy~\cite{Groth_PRB_2016,Dornheim_PRB_2016,groth_prl,Brown_PRL_2013,ksdt,status} and chemical potential~\cite{ChemPot}, momentum distribution~\cite{MILITZER201913,Dornheim_PRB_nk_2021,Dornheim_PRE_2021,Hunger_PRE_2021}, and static linear density response~\cite{dynamic_folgepaper,dornheim_ML,Dornheim_PRL_2020_ESA,Dornheim_PRB_ESA_2021,dornheim_HEDP,Dornheim_HEDP_2022,Dornheim_PRR_2022,Hou_PRB_2022,groth_jcp}. These efforts also gave rise to a variety of impressive fundamental advances, e.g., on the theory of non-linear response functions~\cite{Mikhailov_PRL,Mikhailov_Annalen,Dornheim_PRL_2020,Dornheim_PRR_2021,Tolias_EPL_2023,Dornheim_JCP_ITCF_2021,Dornheim_JPSJ_2021,Vorberger_JStatPhys_2025,Moldabekov_JCTC_2022,Moldabekov_PRB_2026,svensson2026reweightingestimatorsdensityresponse}, new concepts for the analytical continuation of imaginary-time correlation functions (ITCFs)~\cite{dornheim_dynamic,Hamann_PRB_2020,Filinov_PRB_2023,BENEDIXROBLES2026109904,Chuna_JPA_2025,barnfield2026scaleshapedualnewtonmethod,chuna2025noiselesslimitimprovedpriorlimit} (and extraction of frequency moments~\cite{Dornheim_moments_2023,Dornheim_MRE_2023}), and novel closures for dielectric theories~\cite{tanaka_hnc,Tanaka_CPP_2017,Tolias_JCP_2021,castello2021classical,dornheim_electron_liquid,Tolias_JCP_2023,Tolias_PRB_2024}.

The two-dimensional UEG or 2DEG, which is relevant, e.g., for electronic heating in semiconductors~\cite{Ando_1982} and for heated lattice structures~\cite{Sarma_2004}, has been studied somewhat less intensively, although here, too, highly accurate ground-state QMC results~\cite{Tanatar_1989,Drummond_2DEG_2013,Smith_2DEG_2024} have been available for some time. At finite temperatures, results have so far been mostly limited to dielectric theories~\cite{Schweng_1994,Yurtsever_2003,Khanh_2007,Bhukal_2019,kalkavouras2026dielectricformalism2duniform} and semi-empirical classical mappings~\cite{Perrot_2DEG_2001,Dharmawardana_2DEG_2003,Khanh_2004,Dharmawardana_2DEG_2012}.
Most recently, Kalkavouras \textit{et al.}~\cite{kalkavouras2026dielectricformalism2duniform}
have applied the finite-temperature versions of the dielectric schemes by Singwi-Tosi-Land-Sj\"olander (STLS)~\cite{stls_original,stls,stls2} and the hypernetted chain (HNC) extension first implemented by Tanaka for the 3D UEG~\cite{tanaka_hnc} to the two-dimensional electron gas, which has allowed them to present extensive results for its structural and linear density response properties, as well as its interaction and XC-free energy.
In the same paper, they have used new \emph{ab initio} PIMC results for the static structure factor $S(\mathbf{q})$ and the static linear density response function $\chi(\mathbf{q})$ as a quasi-exact (i.e., exact within the given statistical Monte Carlo error bars) benchmark for the dielectric theories, but without presenting and analyzing the PIMC simulations in appropriate detail. In this work, we aim to fill this gap by presenting a dedicated work with particular focus on our PIMC approach to the 2DEG.
First, we provide practical details about the implemented 2D Ewald sum~\cite{Osychenko20022012} and additional details about the fermion sign problem~\cite{dornheim_sign_problem,troyer} of the 2DEG. Second, we cover a broader range of parameters, including the strongly correlated electron liquid with a Wigner-Seitz radius of $r_s\leq50$ and higher temperatures of up to $\Theta=k_\textnormal{B}T/E_\textnormal{F}=16$.
Third, we complement the analysis of structural properties from Ref.\cite{kalkavouras2026dielectricformalism2duniform} with an in-depth investigation of the imaginary-time density--density correlation function (ITCF) $F(\mathbf{q},\tau)=\braket{\hat{n}(\mathbf{q},\tau)\hat{n}(-\mathbf{q},0)}$~\cite{Dornheim_PTR_2023,Dornheim_MRE_2023} that is famously related to the dynamic structure factor $S(\mathbf{q},\omega)$ via
\begin{eqnarray}\label{eq:Laplace}
    F(\mathbf{q},\tau) = \int_{-\infty}^\infty \textnormal{d}\omega\ S(\mathbf{q},\omega)\ e^{-\tau\omega}\ ,
\end{eqnarray}
and corresponds to the more familiar intermediate scattering function $F(\mathbf{q},t)$, but evaluated at the imaginary time argument $t=-i\hbar\tau$ with $\tau\in[0,\beta]$ and $\beta=1/k_\textnormal{B}T$ the inverse temperature in energy units. The numerical inversion of Eq.~(\ref{eq:Laplace}) to reconstruct $S(\mathbf{q},\omega)$ from quasi-exact PIMC results for $F(\mathbf{q},\tau)$ is the exponentially ill-conditioned \emph{analytic continuation} problem~\cite{JARRELL1996133,chuna2025dualformulationmaximumentropy,chuna2025noiselesslimitimprovedpriorlimit}. Here, we avoid the analytic continuation by following recent works by Dornheim and co-workers~\cite{Dornheim_T_2022,Dornheim_MRE_2023,Dornheim_PTR_2023,Dornheim_moments_2023,Chuna_JCP_2025,gawne2026modelfreeinterpretationxraythomson,kalkavouras2026kineticenergycubicsum}, who suggested to work directly in the imaginary-time domain. In particular, we analyze the relative decay of the ITCF with $\tau$ and find clear signatures of a \emph{roton-type} excitation, which has been reported in previous studies of the 3D electron liquid~\cite{dornheim_dynamic,dynamic_folgepaper,Dornheim_Nature_2022,Dornheim_MRE_2023,koskelo2023shortrange,Panholzer_PRL_2018,Takada_PRB_2016,Filinov_PRB_2023,Chuna_PRB_2025,Chuna_JCP_2025} (and also warm dense hydrogen~\cite{Hamann_PRR_2023}) and also in both experimental and theoretical studies of the dynamic structure factor of two-dimensional helium films~\cite{Godfrin2012}. We note that the roton feature is a direct consequence of electronic exchange--correlation effects, and as such constitutes a particularly revealing testbed for approximate models such as dielectric theories.

We are confident that our results will complement the rich body of literature 
on the archetypal UEG model. For example, the weak dependence of the 2DEG on the temperature for certain parameters might make our results directly relevant for the development and the benchmarking of corresponding exchange--correlation functionals for density functional theory~\cite{PhysRevLett.88.256601} and other applications pertinent to the description of thin films and surfaces. In addition, all PIMC results are freely available online~\cite{repo} and can be used as a rigorous benchmark for the development of new methods and approximations, such as improved closure relations for both static and dynamic local field corrections~\cite{kugler1}. The paper is organized as follows: in Sec.\ref{sec:theory}, we introduce the theoretical background, including an overview of the relevant reduced parameters (\ref{sec:parameters}), the explicit form of the model Hamiltonian and Ewald sum (\ref{sec:Hamiltonian}), the PIMC method (\ref{sec:PIMC}), as well as a concise overview of a few key relations from linear density response theory (\ref{sec:LRT}). Our new simulation results are discussed in Sec.\ref{sec:results} starting with a verification and convergence study (\ref{sec:convergence}), a brief analysis of the system-size dependence (\ref{sec:size_matters_not}) and a short overview of the fermion sign problem (\ref{sec:sign_problem}). In Sec.\ref{sec:structure}, we present our extensive new \emph{ab initio} PIMC results for the static structure factor $S(\mathbf{q})$ and static linear density response $\chi(\mathbf{q})$ at a broad range of densities ($r_s=0.1,\dots,50$) and temperatures ($\Theta=0.5,\dots,16$), and compare them with the most advanced existing dielectric theories for the 2DEG. Sec.\ref{sec:spectral} is dedicated to the analysis of the spectral properties of the 2DEG focusing on the ITCF $F(\mathbf{q},\tau)$, while Sec.\ref{sec:roton} provides an in-depth imaginary-time perspective on the \emph{roton-type} feature of the dynamic structure factor. The paper is concluded with a discussion and an outlook in Sec.\ref{sec:outlook}.

\section{Theory\label{sec:theory}}

We use Hartree atomic units throughout.

\subsection{Parameters\label{sec:parameters}}

We consider a quadratic simulation cell of $L=\sqrt{\pi N}r_s$ length and $\Omega=L^2$ area; the inverse temperature is given by $\beta=r_s^2/\Theta$, with $\Theta=k_\textnormal{B}T/E_\textnormal{F}$ the reduced temperature, $E_\textnormal{F}=q_\textnormal{F}^2/2$ the Fermi energy and $q_\textnormal{F}=\sqrt{2}/r_s$ the Fermi wavenumber. Similar to the 3D UEG~\cite{quantum_theory}, the 2DEG attains the limit of ideal (i.e., non-interacting) Fermi gas for high densities ($r_s\to0$). Conversely, the system becomes a strongly coupled liquid and eventually a Wigner crystal~\cite{PhysRevLett.88.256601,10.1071/PH960161} for large Wigner-Seitz radii, although the precise location of these transitions depends on the reduced temperature $\Theta$ and remains uncertain even at zero temperature $\Theta-0$. The reduced temperature quantifies the degree of quantum degeneracy, with $\Theta\ll1$ and $\Theta\gg1$ corresponding to fully degenerate and semi-classical systems, respectively. In this work, we focus on the intermediate regime of $\Theta\sim1$, where quantum degeneracy and delocalization effects play an important but not necessarily dominant role. Throughout this work, we limit ourselves to the paramagnetic UEG with $N^\uparrow = N^\downarrow = N/2$.

\subsection{Hamiltonian\label{sec:Hamiltonian}}

We consider the case of a bare Coulomb pair potential $\phi_\textnormal{C}(\mathbf{r}_1,\mathbf{r}_2)=1/|\mathbf{r}_1-\mathbf{r}_2|$ as well as its regularized Fourier transform $v(\mathbf{q})=2\pi/|\mathbf{q}|$. Following the notation of Osynchenko \emph{et al.}~\cite{Osychenko20022012}, we express the full Hamiltonian that includes the Ewald sum over the infinite array of periodic images as
\begin{eqnarray}
    \hat{H} = -\frac{1}{2} \sum_{l=1}^N \nabla_l^2 + \sum_{l<k}^N W_\textnormal{E}(\hat{\mathbf{r}}_l,\hat{\mathbf{r}}_k) + \frac{N}{2}\xi_\textnormal{M}\ ,
\end{eqnarray}
with the definition of the 2D Ewald sum
\begin{widetext}
\begin{eqnarray}\label{eq:2DEWALD}
    W_\textnormal{E}(\mathbf{r}_1,\mathbf{r}_2) = \sum_{\mathbf{n}} \frac{ \textnormal{erfc}\left(\alpha|\mathbf{n}+(\mathbf{r}_1-\mathbf{r}_2)L^{-1}|\right) }{ |\mathbf{n}L + (\mathbf{r}_1-\mathbf{r}_2)| }
    +\sum_{\mathbf{n}\neq\mathbf{0}}\left\{
    \textnormal{cos}\left( 2\pi L^{-1} \mathbf{n}\cdot (\mathbf{r}_1-\mathbf{r}_2) \right)
    \frac{\textnormal{erfc}\left( \frac{\pi n}{\alpha} \right)}{nL}
    \right\}
    - 2 \frac{\sqrt{\pi}}{\alpha L }
\end{eqnarray}
\end{widetext}
and the (dimensionless) Madelung constant
\begin{eqnarray}\label{eq:2DMADELUNG}
    \xi_\textnormal{M} = \sum_{\mathbf{n}\neq\mathbf{0}} \left\{
\frac{ \textnormal{erfc}\left(\alpha n\right) + \textnormal{erfc}\left(\frac{\pi n}{\alpha}\right) }{n}
    \right\}
    - \frac{2\sqrt{\pi}}{\alpha} - \frac{2\alpha}{\sqrt{\pi}}\ 
\end{eqnarray}
that takes into account the interaction between a charge and its own infinite array of background images. The Madelung constant can be conveniently parametrized as
\begin{equation}
\frac{N}{2}\xi_\textnormal{M} \approx -1.10024442048(1) N^{1/2} / r_s\ .
\end{equation}
Note that the full Eqs.~(\ref{eq:2DEWALD}) and (\ref{eq:2DMADELUNG}) do not depend on the Ewald parameter $\alpha$; the latter is thus chosen to optimize the simultaneous convergence of the real and reciprocal space sums in Eq.~(\ref{eq:2DEWALD}). We find $\alpha\approx3.5$ to be a good empirical choice for the parameters considered in this work.

\begin{figure}
    \centering
    \includegraphics[width=0.445\textwidth]{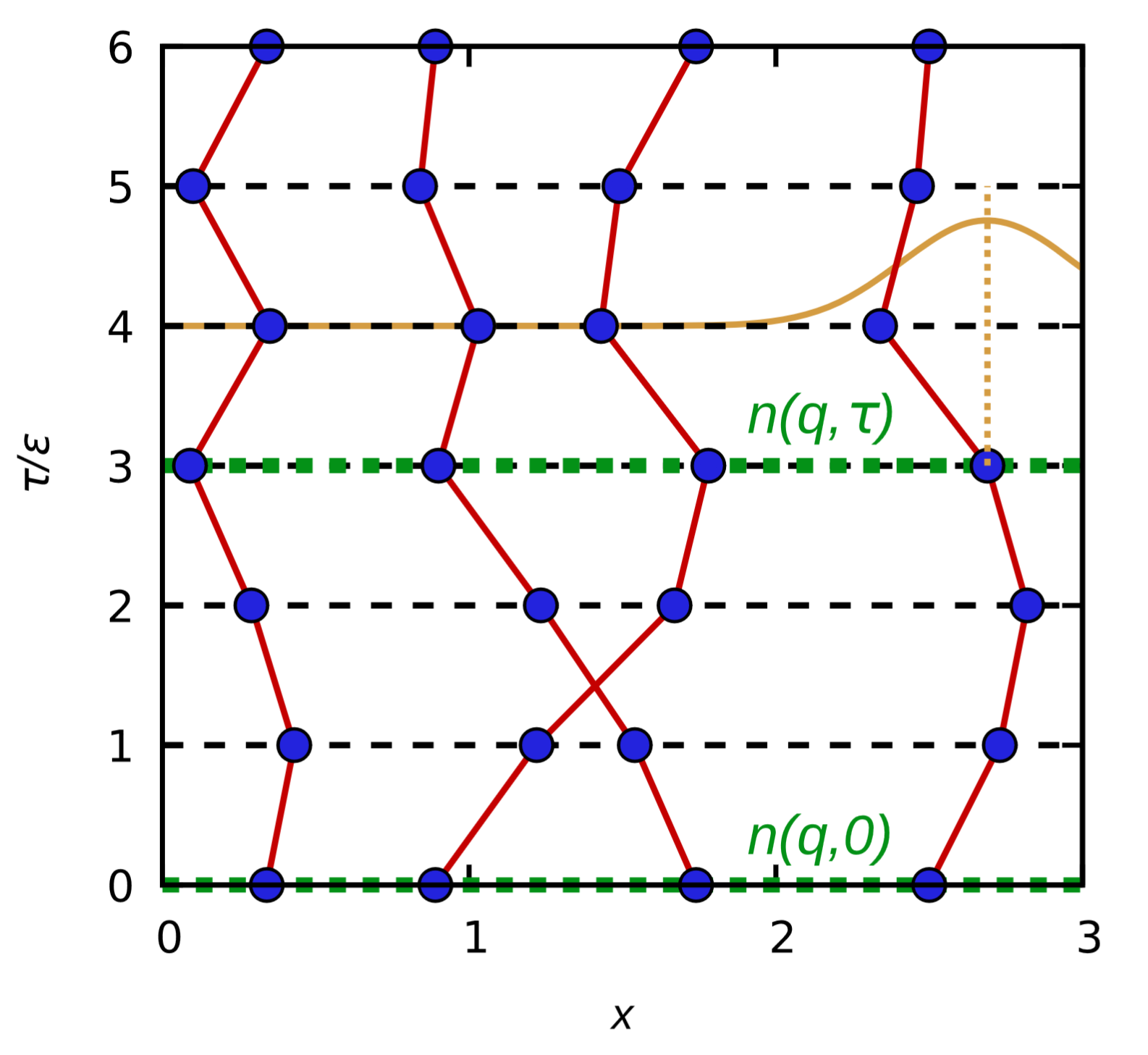}
    \caption{Schematic illustration of a PIMC configuration of $N=4$ electrons in the $\tau$-$x$-plane. The yellow Gaussian on the right illustrates the free (kinetic) imaginary-time propagator [Eq.~(\ref{eq:ideal_propagator})] and the two horizontal dashed green lines illustrate the estimation of the ITCF $F(\mathbf{q},\tau)$ [Eq.~(\ref{eq:define_ITCF})] within PIMC. Adopted from Ref.\cite{Dornheim_MRE_2023} with the permission of the authors.}
    \label{fig:scheme}
\end{figure}

\subsection{Path integral Monte Carlo\label{sec:PIMC}}

Having originally been developed for the accurate description of ultracold quantum liquids~\cite{Fosdick_PR_1966,cep}, the \emph{ab initio} PIMC method has emerged as one of the most powerful techniques for the simulation of interacting quantum many-body systems in thermal equilibrium. Implementations have been demonstrated for different thermodynamic and generalized ensembles~\cite{Marienhagen_JCP_2024,Marienhagen_JCP_2025,boninsegni1,hamann2026abinitiopathintegralmonte,Dornheim_PRB_nk_2021,mezza}, and the development of new update schemes remains an active topic of research~\cite{zhao2026exchangemontecarlocontinuousspace}.

Here, we consider the canonical (i.e., inverse temperature $\beta$, number density $n=N/\Omega$ and area $\Omega$ are fixed) ensemble with the corresponding partition function
\begin{eqnarray}\label{eq:Z}
    Z = \frac{1}{N^\uparrow! N^\downarrow !} \sum_{\sigma_{\uparrow\downarrow}\in S_{N^\uparrow,N^\downarrow}} \textnormal{sgn}\left(\sigma_{\uparrow\downarrow}\right) \int \textnormal{d}\mathbf{R}\ \bra{\mathbf{R}} e^{-\beta\hat{H}} \ket{\hat{\pi}_{\sigma_{\uparrow\downarrow}}\mathbf{R}}
\end{eqnarray}
expressed in coordinate representation, where the meta variable $\mathbf{R}=(\mathbf{r}_1,\dots,\mathbf{r}_N)^T$ contains the coordinates of all $N$ electrons.
Since electrons are fermions, the partition function has to be properly anti-symmetrized by the sum over all possible permutations $\sigma_{\uparrow\downarrow}$, with $\hat{\pi}_{\sigma_{\uparrow\downarrow}}$ the corresponding coordinate exchange operator, and where the sign function $\textnormal{sgn}\left(\sigma_{\uparrow\downarrow}\right)$ is positive (negative) for an even (odd) number of pair exchanges.

The key problem with Eq.~(\ref{eq:Z}) is that the matrix elements of the canonical density operator $\hat\rho=e^{-\beta\hat{H}}$ cannot be readily evaluated as the kinetic and potential contributions to the Hamiltonian $\hat{H}=\hat K + \hat V$ do not commute.
To overcome this obstacle, we apply the usual exact $\hat\rho$ semi-group property to express the density operator at inverse temperature $\beta$ as an integral over $P-1$ density matrices at $P$ times the original temperature, $\epsilon=\beta/P$. In the limit of $P\to\infty$, we can safely employ the primitive factorization $e^{-\epsilon\hat{H}}\approx e^{-\epsilon\hat{K}}e^{-\epsilon\hat{V}}$, which is justified by the famous Trotter formula~\cite{trotter} and asymptotically converges as $\mathcal{O}\left(P^{-2}\right)$. For completeness, we note that higher-order factorization schemes have been explored in the literature~\cite{Chin_PRE_2015,sakkos_JCP_2009,Dornheim_NJP_2015,Dornheim_CPP_2019,Zillich_JCP_2010,cep}, but are not required here.

The final result for the PIMC partition function can be expressed symbolically in the compact form
\begin{eqnarray}\label{eq:Z_compact}
    Z = \sumint \textnormal{d}\mathbf{X}\ W(\mathbf{X})\ ,
\end{eqnarray}
where $\mathbf{X}=(\mathbf{R}_0,\dots,\mathbf{R}_{P-1})^T$ contains $P$ separate sets of particle coordinates over which one has to integrate, and where $\sumint \textnormal{d}\mathbf{X}$ also contains the sum over all permutations of particle coordinates that is more explicit in Eq.~(\ref{eq:Z}). The weight function $W(\mathbf{X})$ contains both kinetic and interaction terms and can be readily evaluated.

The guiding paradigm behind the PIMC approach is the celebrated classical isomorphism~\cite{Chandler_JCP_1981} making use of the formal equivalence between the density operator at an inverse temperature $\tau$ and the time evolution operator evaluated at the imaginary time argument $t=-i\tau$. As a consequence, we can map any quantum many-body system in thermal equilibrium onto an effectively classical ensemble of interacting ring polymers---the eponymous \emph{paths} of PIMC. This is illustrated in Fig.\ref{fig:scheme}, where we show a schematic configuration of $N=4$ electrons with $P=6$ high-temperature factors. Each electron is represented by a trajectory along the imaginary time $\tau\in[0,\beta]$, with $P$ discrete imaginary time slices of length $\Delta\tau=\epsilon$; this can be understood as a Gaussian diffusion process along $\tau$ that is primarily guided by the kinetic (and, thus, non-interacting) contribution to the full density matrix~\cite{Dornheim_PTR_2023}, which, for a pair of particles, reads as:
\begin{eqnarray}\label{eq:ideal_propagator}
    \rho_0(\mathbf{r}_1,\mathbf{r}_2;\Delta\tau) = \frac{1}{\lambda_{\Delta\tau}^2} \sum_{\mathbf{n}}\textnormal{exp}\left(
-\frac{\pi}{\lambda_{\Delta\tau}^2} (\mathbf{r}_1 - \mathbf{r}_2 + \mathbf{n}L)^2
    \right)\ ,
\end{eqnarray}
which is a Gaussian with variance $\sigma=\sqrt{\Delta\tau}$, where $\lambda_{\Delta\tau}=\sqrt{2\pi\Delta\tau}$ is the reduced thermal wavelength. We note that the potential energy operator simply contributes in the form of an effective Boltzmann factor of the form $e^{-\epsilon V(\mathbf{R})}$ on each time slice.
For high temperatures, the Gaussian propagators (see the dark yellow curve in the top right corner of Fig.~\ref{fig:scheme}) in Eq.~(\ref{eq:ideal_propagator}) narrow and the paths become less extended. In the limit of very high temperatures, the paths will resemble straight lines, corresponding to classical point particles. Conversely, the Gaussians become wider for lower temperatures, leading to more spread out trajectories that, in turn, correspond a higher degree of quantum delocalization.

The key idea of the PIMC method is to employ modern implementations of the acclaimed Metropolis algorithm~\cite{metropolis} to generate a Markov chain of configurations $\{\mathbf{X}_j\}$ that are distributed according to the probability $P(\mathbf{X})=W(\mathbf{X})/Z$. We note that this set must also include all possible permutations of particle coordinates, which, in the imaginary-time path integral picture, manifest as \emph{permutation cycles}~\cite{Dornheim_permutation_cycles}, i.e., trajectories with more than a single particle; see the center of Fig.~\ref{fig:scheme} for an example. For bosons (and also for hypothetical distinguishable quantum particles, sometimes denoted as \emph{boltzmannons} in the literature), all contributions to Eq.~(\ref{eq:Z_compact}) are strictly non-negative, and modern implementations allow for quasi-exact simulations of up to $N\sim10^4$ quantum particles~\cite{boninsegni1,boninsegni2}. For fermions, such as the electrons that are being considered in the present work, the configuration weight $W(\mathbf{X})$ can be both positive and negative depending on the number of pair exchanges; this precludes the interpretation of $P(\mathbf{X})$ as a proper probability distribution suitable for straightforward Monte Carlo sampling.
As a practical workaround, we consider the modified partition function
\begin{eqnarray}\label{eq:Z_modified}
    Z' = \sumint\textnormal{d}\mathbf{X}\ |W(\mathbf{X})|\ ,
\end{eqnarray}
and sample path configurations according to the corresponding probability $P'(\mathbf{X})$.
The fermionic expectation value of any observable of interest $\hat{\gamma}$ is then given by~\cite{dornheim_sign_problem}
\begin{eqnarray}\label{eq:ratio}
    \braket{\hat\gamma} = \frac{\braket{\hat\gamma\hat S}'}{\braket{\hat S}'}\ ,
\end{eqnarray}
with $S(\mathbf{X})=|W(\mathbf{X})|^{-1}W(\mathbf{X})$ the sign associated with a configuration $\mathbf{X}$. The denominator of Eq.~(\ref{eq:ratio}) is commonly known as the average sign $S$ in the quantum Monte Carlo community and directly quantifies the amount of cancellations of positive and negative contributions. In practice, simulations are feasible for $S\gtrsim0.01-0.001$, since the statistical error of any Monte Carlo average scales, to first order, as $\sigma_\gamma\sim1/S$~\cite{hatano1994data}, which can only be compensated by increasing the number of Monte Carlo samples $N_\textnormal{MC}$ as $1/\sqrt{N_\textnormal{MC}}$. At the same time, the average sign decays exponentially upon increasing the system size or decreasing the temperature. This de-facto exponential wall is known as the \emph{fermion sign problem}~\cite{troyer,dornheim_sign_problem,Loh_PRB_1990}, and constitutes the main limitation of our simulations. The potential application of novel methods that are capable of dealing with the sign problem somewhat more efficiently~\cite{Hirshberg_JCP_2020,Dornheim_Bogoliubov_2020,Xiong_JCP_2022,Dornheim_JCP_xi_2023,Dornheim_NatComm_2025,Dornheim_PRR_2026} is discussed in the Outlook~\ref{sec:outlook}.

\subsection{Linear response theory and the ITCF\label{sec:LRT}}

The complete wavenumber- and frequency-resolved information about the density response of a given quantum many-body system to an external harmonic perturbation is encoded into the dynamic density response function $\chi(\mathbf{q},\omega)$~\cite{quantum_theory}. On the one hand, $\chi(\mathbf{q},\omega)$ is fully determined by the equilibrium properties of the unperturbed system. On the other hand, complete knowledge of $\chi(\mathbf{q},\omega)$ is, in principle, also sufficient to determine all equilibrium properties of the system of interest, making it the ongoing focus of theory and experiment alike~\cite{Dornheim_review}.
For example, the fluctuation--dissipation theorem~\cite{quantum_theory}
\begin{eqnarray}\label{eq:FDT}
S(\mathbf{q},\omega) = - \frac{\textnormal{Im}\chi(\mathbf{q},\omega)}{\pi n (1-e^{-\beta\omega})}\ ,
\end{eqnarray}
relates the dynamic structure factor---the key observable in x-ray Thomson scattering~\cite{siegfried_review} and electron energy loss spectroscopy~\cite{8wwb-lxk2} experiments---to the density response. Without loss of generality, the latter can be conveniently expressed as~\cite{kugler1}
\begin{eqnarray}\label{eq:define_LFC}
    \chi(\mathbf{q},\omega) = \frac{\chi_0(\mathbf{q},\omega)}{1-v(\mathbf{q})\left[1-G(\mathbf{q},\omega)\right]\chi_0(\mathbf{q},\omega)}\ ,
\end{eqnarray}
with $v(q) = 2\pi/q$ the Fourier transform of the Coulomb pair potential.
For the 2DEG, one usually uses as a reference function $\chi_0(\mathbf{q},\omega)$ the density response of the ideal (i.e., non-interacting) 2D electron gas~\cite{quantum_theory}. In that case, the dynamic local field correction $G(\mathbf{q},\omega)$ contains the full information about electronic exchange--correlation effects, and setting $G(\mathbf{q},\omega)\equiv0$ in Eq.~(\ref{eq:define_LFC}) corresponds to the ubiquitous \emph{random phase approximation} (RPA), which treats the electronic density response on the mean-field level. In addition, we note that the finite temperature implementations of the STLS and HNC schemes for the 2DEG presented in Ref.\cite{kalkavouras2026dielectricformalism2duniform} completely neglect the frequency dependence of the local field correction, i.e., $G(\mathbf{q},\omega)\equiv \overline{G}(\mathbf{q})$, which has been shown to cause some spurious behavior in the 3D case~\cite{Dornheim_PRL_2020_ESA,stls}.

\begin{figure*}
    \centering
    \includegraphics[width=0.445\textwidth]{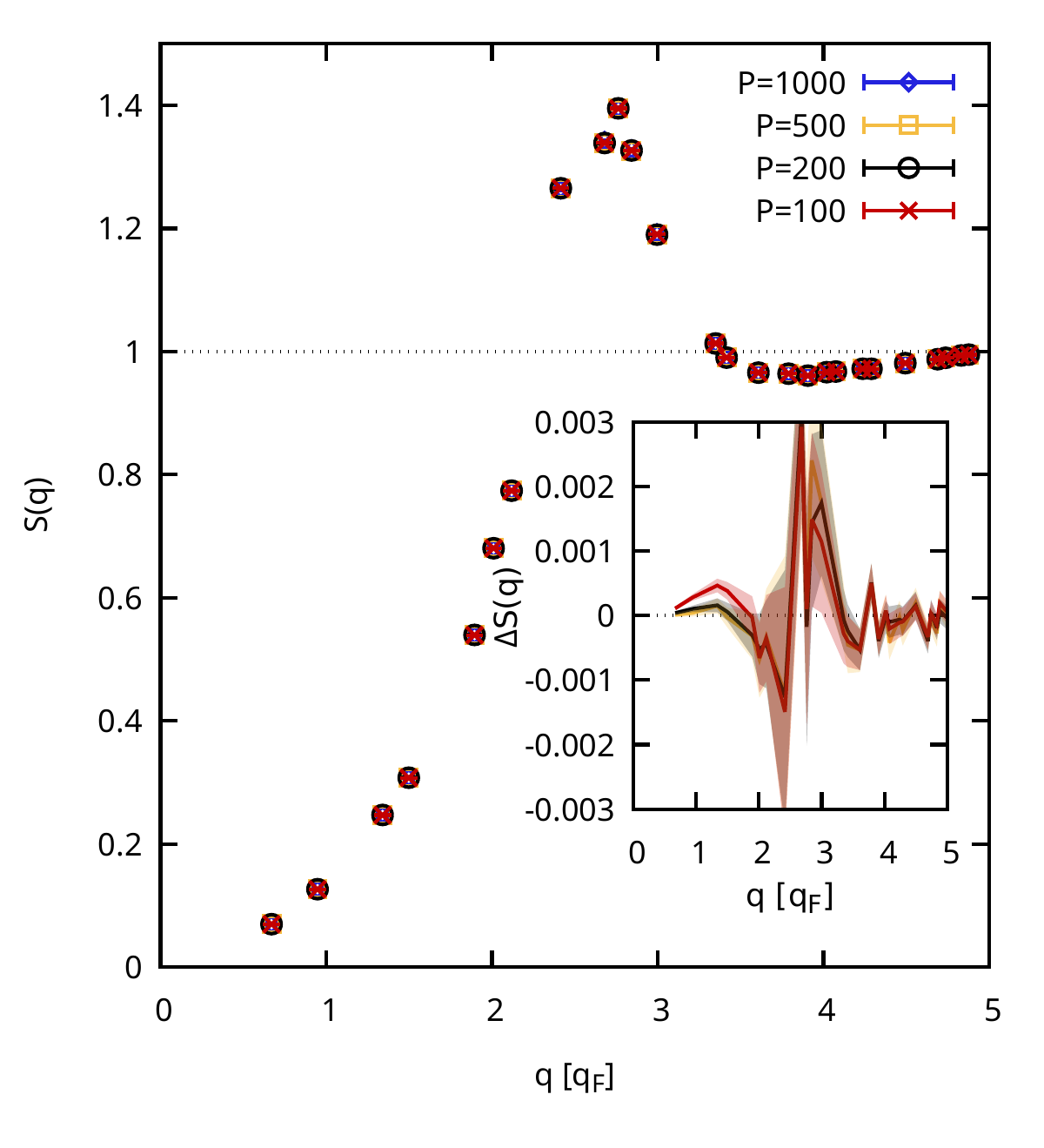} \includegraphics[width=0.445\textwidth]{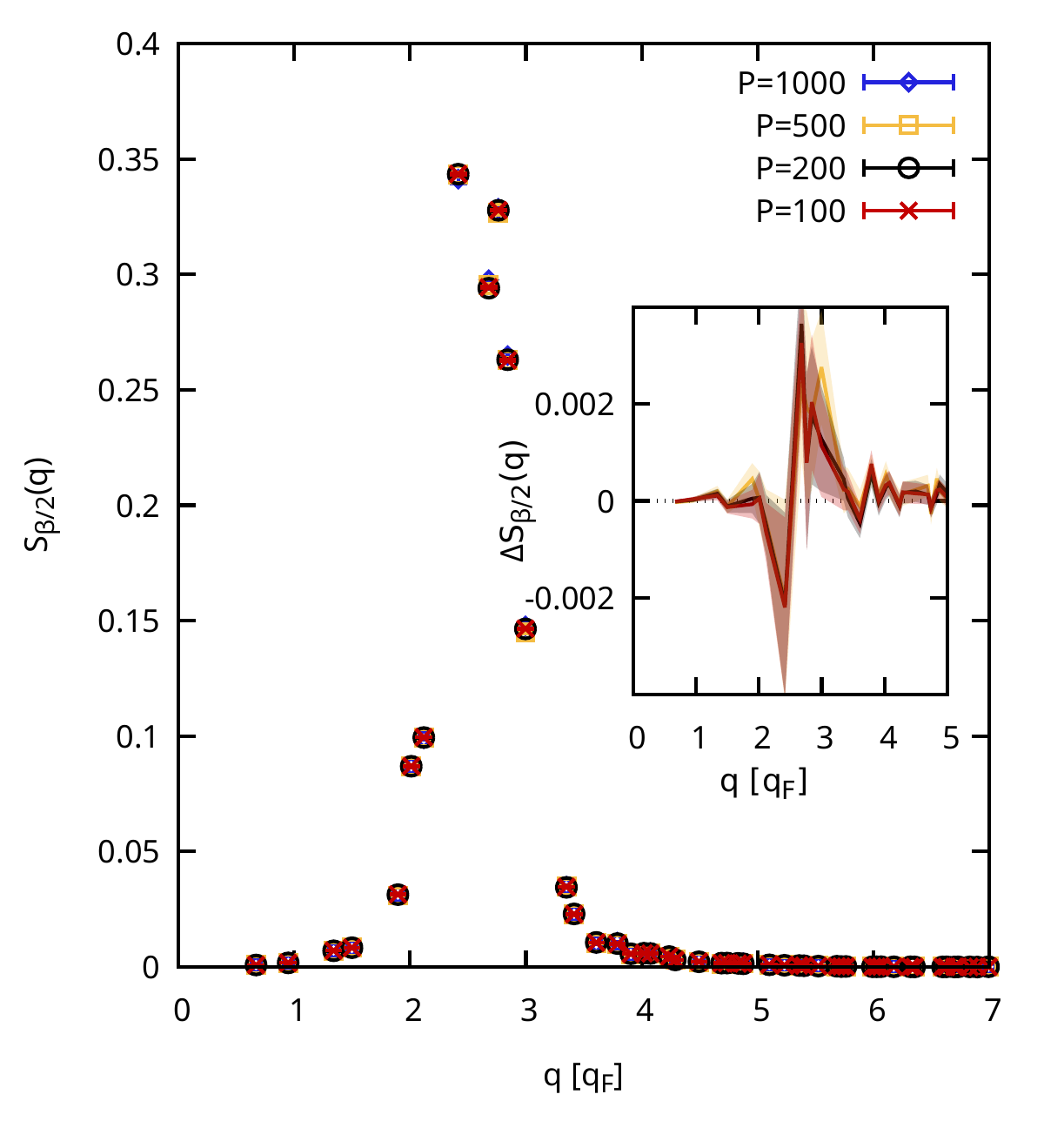}
    \caption{The PIMC convergence with the number of imaginary-time propagators $P$ for the 2DEG at $r_s=30$ and $\Theta=0.5$ with $N=14$. Left: static structure factor $S(\mathbf{q})$ for $P=1000$ (blue diamonds), $P=500$ (yellow squares), $P=200$ (black circles) and $P=100$ (red crosses). The inset shows the differences with the reference results for $P=1000$. Right: same but for the thermal structure factor $S_{\beta/2}(\mathbf{q}) = F(\mathbf{q},\beta/2)$.}
    \label{fig:convergence}
\end{figure*}

There are many ways to study density response properties with the PIMC method. In the static limit $\omega\to0$, one can use the \emph{direct perturbation approach}~\cite{moroni,moroni2,bowen2,dornheim_pre,groth_jcp,Dornheim_PRL_2020} and determine the linear (and non-linear~\cite{Dornheim_PRL_2020,Dornheim_PRR_2021}) density response from simulation results for the perturbed system. Although formally exact, this approach requires a potentially large number of simulations for different perturbation wavenumbers and amplitudes, thus generally precluding extensive parameter scans.
Very recently, Svensson \textit{et al.}~\cite{svensson2026reweightingestimatorsdensityresponse} have explored a re-weighting implementation of this idea, which is capable of capturing the full (both linear and non-linear) response from a single simulation of the unperturbed system.
An elegant alternative to the direct perturbation approach is given by the well-known imaginary-time version of the fluctuation--dissipation theorem~\cite{bowen2,Dornheim_MRE_2023}
\begin{eqnarray}\label{eq:static_chi}
   \chi(\mathbf{q}) \equiv \chi(\mathbf{q},0) = - n\int_0^\beta\textnormal{d}\tau\ F(\mathbf{q},\tau)\ ,
\end{eqnarray}
relating the static linear density response to the area under the ITCF $F(\mathbf{q},\tau)$, which has already been introduced in the context of Eq.~(\ref{eq:Laplace}) above. Within our PIMC simulations, we directly estimate $F(\mathbf{q},\tau)$ from its definition 
\begin{eqnarray}\label{eq:define_ITCF}
    F(\mathbf{q},\tau) = \braket{\hat{n}(\mathbf{q},0)\hat{n}(-\mathbf{q},\tau)}\ ,
\end{eqnarray}
i.e., by correlating two single-particle density operators in reciprocal space at an imaginary-time difference $\tau$; this is illustrated by the two dashed green horizontal lines in Fig.~\ref{fig:scheme} above. For that reason, we can access the ITCF (and, in principle, a number of other imaginary-time correlation functions~\cite{boninsegni1,hamann2026abinitiopathintegralmonte,Filinov_PRA_2012,Dornheim_JCP_ITCF_2021,Reichman_Rabani_JCP_2002,efremkin2026computationthermalconductivitybased}) on the discrete $\tau$-grid that follows from the $P$ imaginary-time slices introduced in Sec.~\ref{sec:PIMC}. Numerically evaluating the integral in Eq.~(\ref{eq:static_chi}) works very well for a moderate grid of $P=200$, except in the limit of very large wavenumbers, $q\gg q_\textnormal{F}$.

The study of the frequency dependence of the dynamic density response~\cite{Hamann_PRB_2020} or dynamic structure factor would require one to numerically invert the two-sided Laplace transform given in Eq.~(\ref{eq:Laplace}). As stated above, this is the exponentially ill-posed analytic continuation problem, which we will not attempt to tackle in the present work.
Nevertheless, our extensive PIMC results for the ITCF itself are already sufficient to give us a gamut of physical insights into the dynamic properties of the studied system.
Indeed, the formal uniqueness of the two-sided Laplace transform directly implies that $F(\mathbf{q},\tau)$ and $S(\mathbf{q},\omega)$ must at least in principle contain exactly the same information, albeit in different representations~\cite{Dornheim_MRE_2023,Chuna_JCP_2025}. A case in point is given by the frequency moments of the dynamic structure factor
\begin{eqnarray}\label{eq:define_moments}
    M_S^{(l)}(\mathbf{q}) &=& \int_{-\infty}^\infty \textnormal{d}\omega\ S(\mathbf{q},\omega)\ \omega^l \\
    &=& (-1)^l \frac{\partial^l}{\partial\tau^l}F(\mathbf{q},\tau)\Big|_{\tau=0}\ ,
\end{eqnarray}
given by the derivatives of the ITCF around $\tau=0$. The straightforward estimation of these derivatives is rendered impractical by the equidistant $\tau$-grid in PIMC. To circumnavigate this obstacle, Dornheim \textit{et al.}~\cite{Dornheim_moments_2023} have suggested to consider the Taylor expansion around $\tau=0$,
\begin{eqnarray}\label{eq:Taylor}
    F(\mathbf{q},\tau) = \sum_{l=0}^\infty \frac{1}{l!} \tau^l \frac{\partial^l}{\partial\tau^l}F(\mathbf{q},\tau)\Big|_{\tau=0} = \sum_{l=0}^\infty c_l(\mathbf{q})\tau^l\ .
\end{eqnarray}
Fitting a canonical polynomial to the PIMC results for the ITCF then allows to obtain the desired frequency moments from the fit coefficients via
$M_S^{(l)}(\mathbf{q}) = (-1)^l l! c_l(\mathbf{q})$. Other physics considerations of the ITCF are introduced in their respective appropriate context within Sec.\ref{sec:results}.

\section{Results\label{sec:results}}

All PIMC results presented in this work have been obtained using the open-source \texttt{ISHTAR} code~\cite{ISHTAR}, and are freely available in an online repository~\cite{repo}.

\subsection{Verification and convergence\label{sec:convergence}}

\begin{figure}
    \centering
    \includegraphics[width=0.445\textwidth]{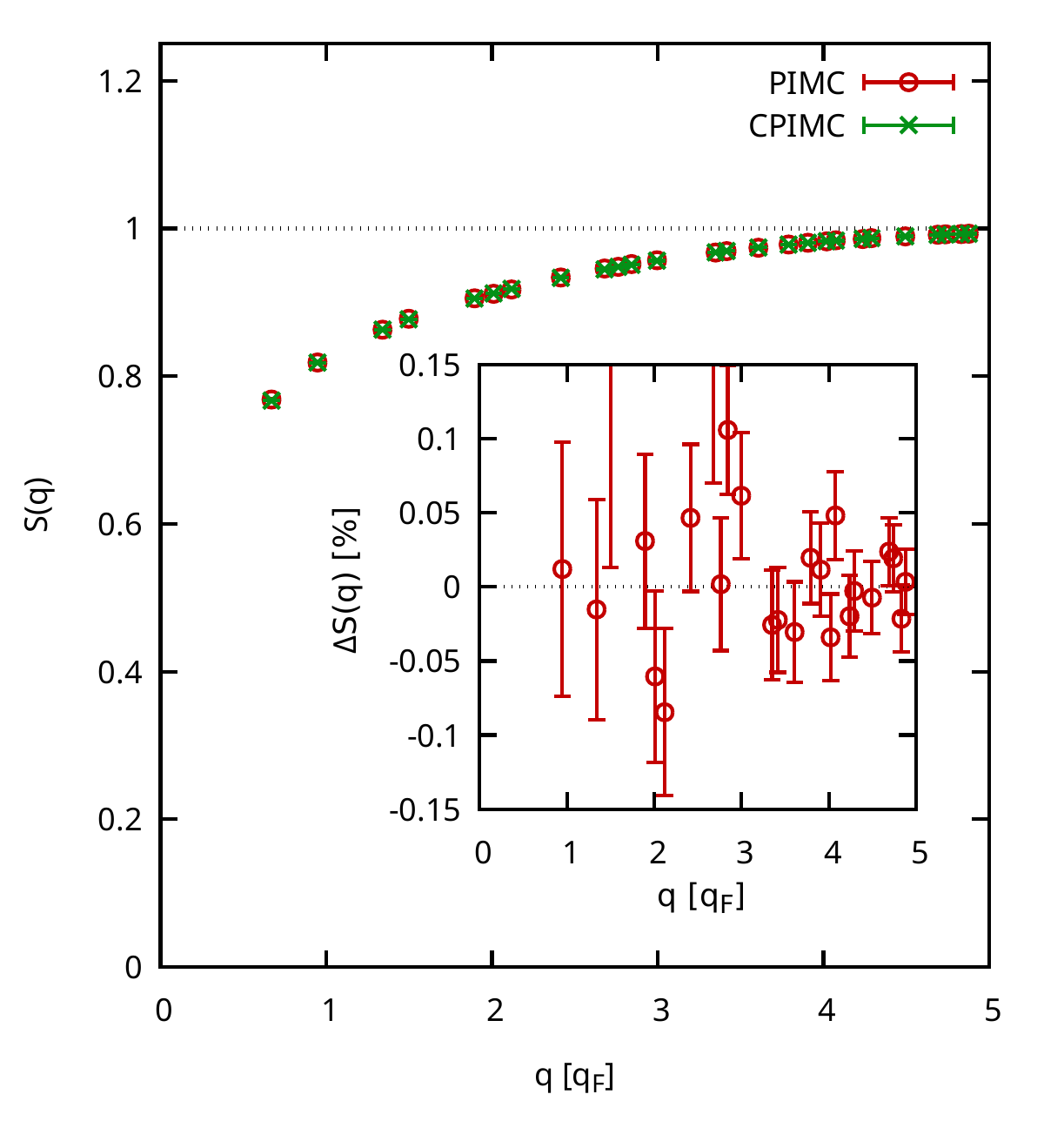}
    \caption{The static structure factor $S(\mathbf{q})$ of the 2DEG at $r_s=0.5$ and $\Theta=4$, with $N=14$ electrons. Red circles: standard PIMC results for $P=200$; green crosses: configuration PIMC results~\cite{cpimc.jl}. The inset shows the relative difference between the two data sets. %\textcolor{red}{$S=0.3405(3)$}
    }
    \label{fig:cpimc}
\end{figure}

We begin with a dedicated investigation of the factorization error for the most challenging (i.e., lowest temperature) conditions considered in the present work. In Fig.~\ref{fig:convergence}, we show PIMC simulation results for the 2DEG with $N=14$ electrons at $r_s=30$ and $\Theta=0.5$, where the blue diamonds, yellow squares, black circles and red crosses correspond to $P=1000$, $P=500$, $P=200$, and $P=100$, respectively. The left panel illustrates the static structure factor $S(\mathbf{q})$, which exhibits a liquid-like behavior evident, e.g., from the first peak exceeding a value of $1.2$~\cite{Kundu_POP_2014}. % Blame Zhandos, not me :). Fucking Zhandos
Importantly, no propagator error can be discerned with the naked eye over the entire depicted range of wavenumbers. The inset shows the deviations from the $P=1000$ reference data set for the other $P$; differences are very small and barely significant even for $P=100$. The right panel of Fig.~\ref{fig:convergence} shows the same information for the thermal structure factor, $S_{\beta/2}(\mathbf{q})=F(\mathbf{q},\beta/2)$. Let us postpone the discussion of its physical interpretation to a later section and exclusively focus on the convergence with $P$. A negligible $P$-dependence is found for all $q$ and it is concluded that $P=200$ constitutes an appropriate choice for all simulations presented throughout this work.

To verify the 2D Ewald sum [Eq.~(\ref{eq:2DEWALD})] implementation into the \texttt{ISHTAR} code~\cite{ISHTAR}, we compare with independent configuration PIMC (CPIMC) results~\cite{cpimc.jl} in Fig.\ref{fig:cpimc} for the $S(\mathbf{q})$, which we expect to be one of the observables most sensitive to the pair potential~\cite{lj9c-bh48}. We note that CPIMC is particularly efficient and highly accurate for high densities and weak to moderate coupling strengths. We, thus, compare results for $r_s=0.5$ and $\Theta=4$ with $N=14$ electrons; the red circles and green crosses correspond to our PIMC implementation and the CPIMC reference data, respectively. We find excellent agreement for all $q$, as expected. This can be seen particularly well in the inset that shows the relative deviations between the two data sets, with an overall agreement of $\lesssim0.1\%$ and no systematic trends.

\begin{figure}
    \centering
    \includegraphics[width=0.445\textwidth]{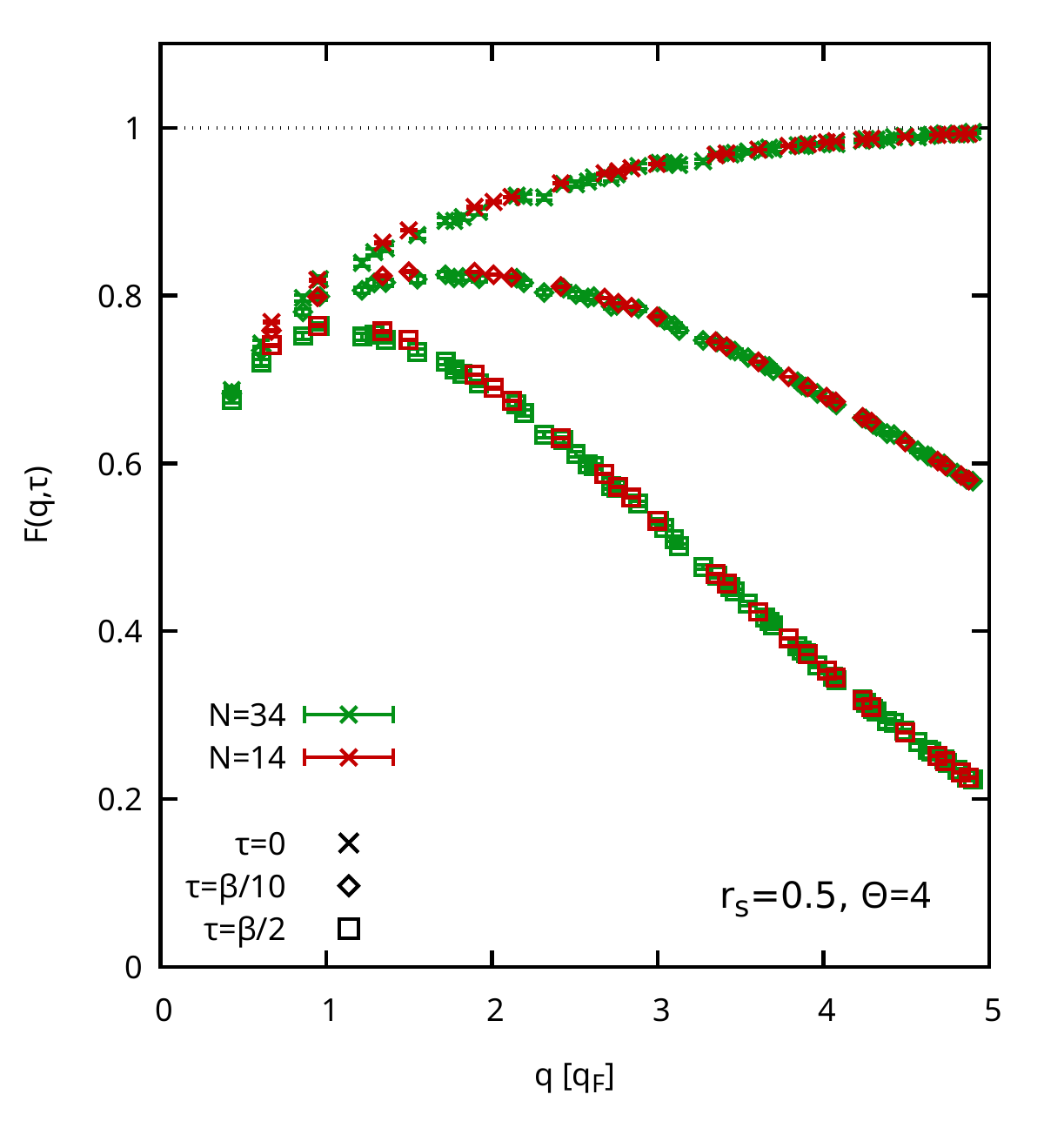}
    \caption{Wavenumber dependence of different ITCF $F(\mathbf{q},\tau)$ $\tau-$slices for $\tau=0$ (crosses), $\tau=\beta/10$ (diamonds) and $\tau=\beta/2$ (squares). PIMC results for the 2DEG at $r_s=0.5$ and $\Theta=4$. The green and red colors distinguish simulation results with $N=34$ and $N=14$ electrons, respectively.}
    \label{fig:FiniteSize1}
\end{figure}

\begin{figure}
    \centering
    \includegraphics[width=0.445\textwidth]{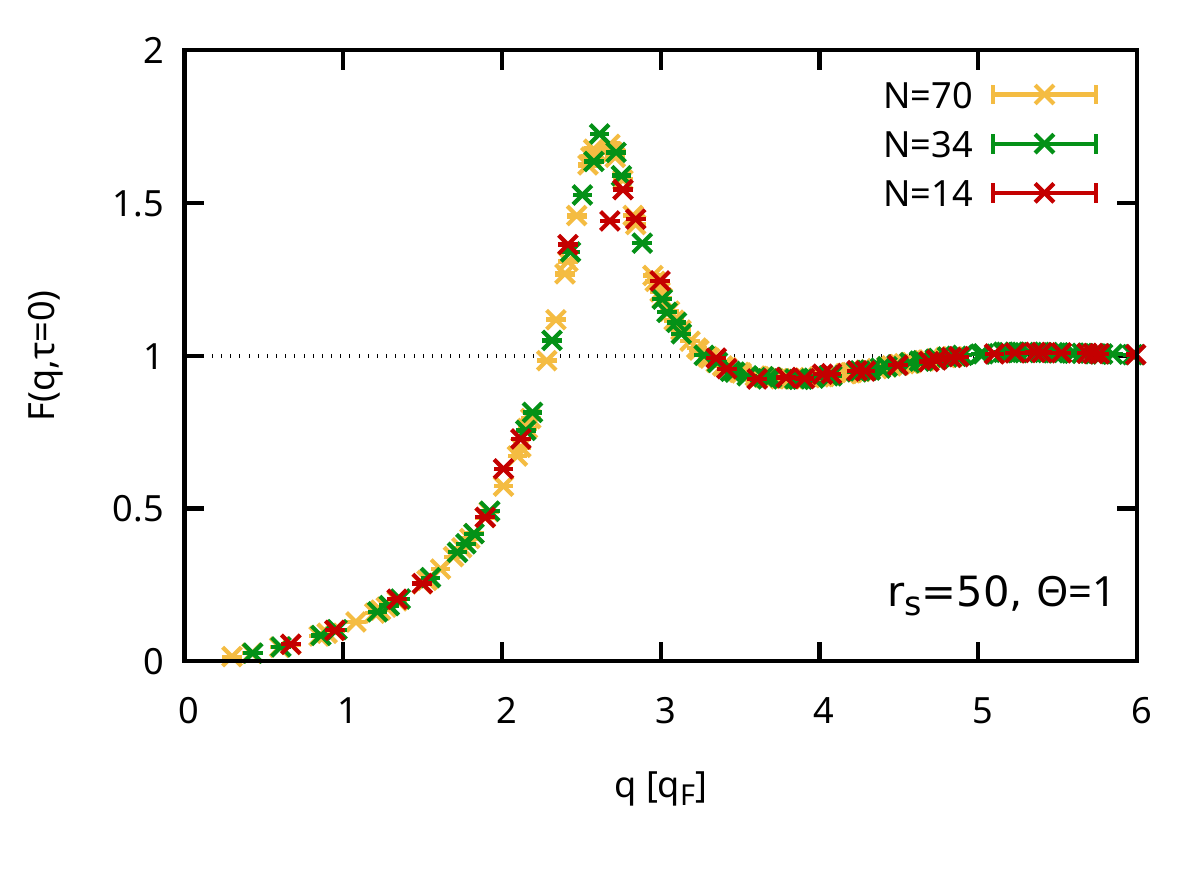}\\\vspace*{-1cm}\includegraphics[width=0.445\textwidth]{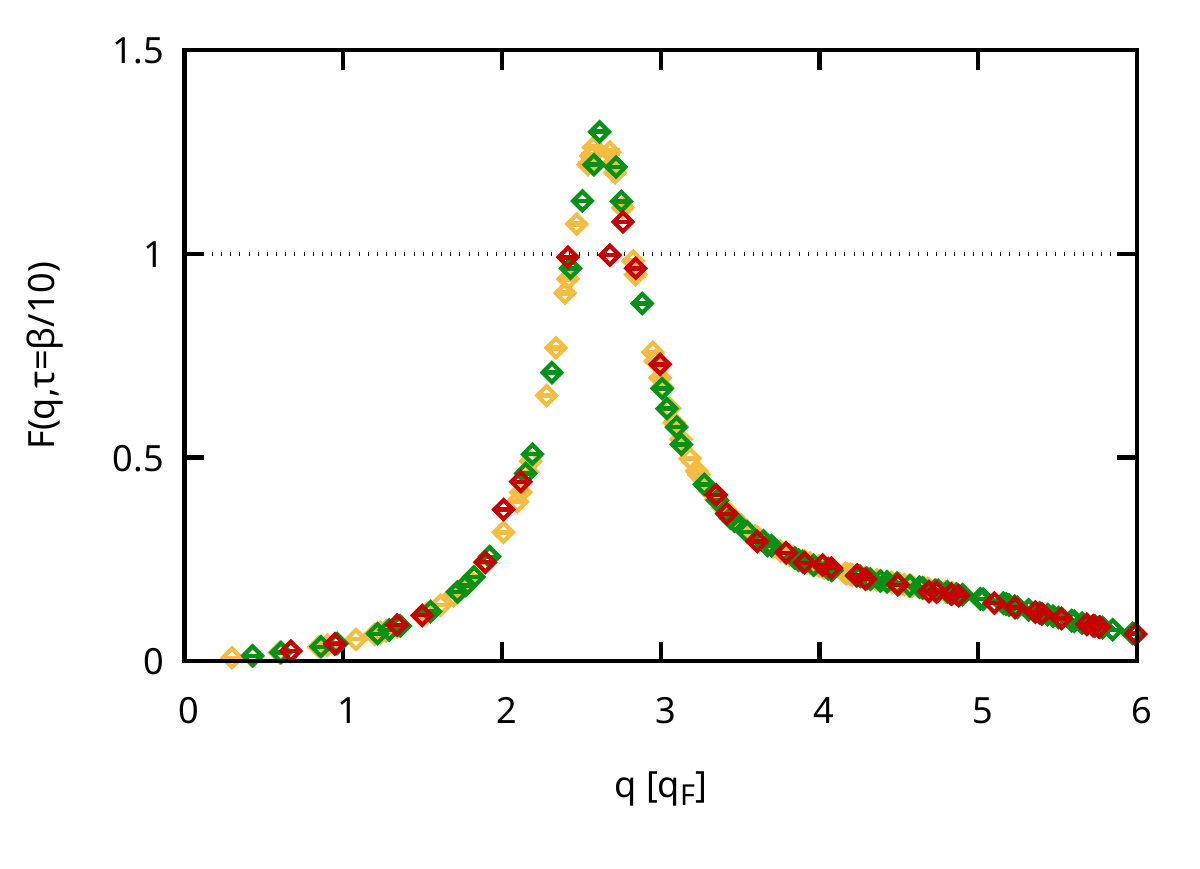}\\\vspace*{-1cm}\includegraphics[width=0.445\textwidth]{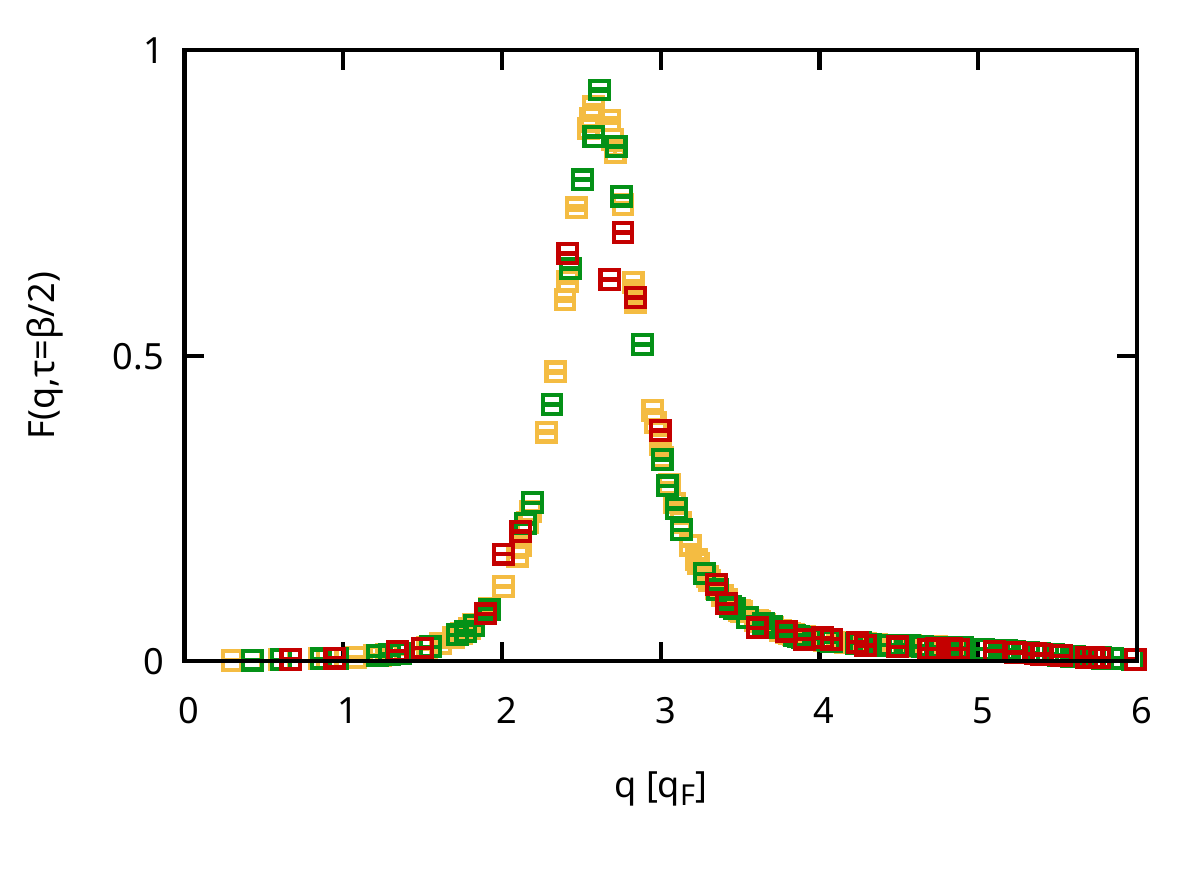}
    \caption{Wavenumber dependence of different ITCF $F(\mathbf{q},\tau)$ $\tau-$slices for $\tau=0$ (crosses), $\tau=\beta/10$ (diamonds) and $\tau=\beta/2$ (squares). PIMC results for the 2DEG at $r_s=50$ and $\Theta=1$. The yellow, green and red colors distinguish simulation results with $N=70$, $N=34$ and $N=14$ electrons, respectively. 
    }
    \label{fig:FiniteSize2}
\end{figure}

\subsection{Dependence on system size\label{sec:size_matters_not}}

Having verified the correctness of our implementation, we next turn to the dependence of our results on the system size.
For the 3D UEG, there exists an extensive body of literature on finite-size effects~\cite{Chiesa_PRL_2006,Brown_PRL_2013,dornheim_prl,Holzmann_PRB_2016,review,Dornheim_JCP_2021,Dornheim_PRE_2020}, which can be summarized as follows: (i) generally, finite-size effects are small in wavenumber resolved properties such as the static structure factor or the ITCF with the main effect concerning the discrete $\mathbf{q}$-grid;
(ii) finite-size effects in integrated properties such as the interaction energy or exchange--correlation free energy~\cite{dornheim2025direct,dornheim2025eta,dornheim2025fermionic,svensson2025accelerated} stem primarily from the approximation of a continuous $\mathbf{q}$-integral in the thermodynamic limit by a discrete $\mathbf{q}$-sum in the finite system and can be corrected to a large degree; (iii) the first point is least fulfilled in the limits of high density and high temperature or low density and low temperature. In the latter case, this is primarily due to the incipient localization and crystallization, where the emerging medium-range order might be affected by a too small simulation box.

In the present work, we exclusively focus on the structural and spectral properties of the 2DEG and we leave point (ii) for a dedicated future investigation. In order to assess points (i) and (iii),  we initially turn our attention to the high-temperature and high-density limit in Fig.~\ref{fig:FiniteSize1}, where we plot the wavenumber dependence of the ITCF $F(\mathbf{q},\tau)$ for $\tau=0$ (crosses), $\tau=\beta/10$ (diamonds) and $\tau=\beta/2$ (squares). The red and green colors distinguish results for $N=14$ and $N=34$ electrons, respectively.
Evidently, the main effect of the system size is indeed given by the differing $\mathbf{q}$-grids, whereas hardly any differences can be resolved between the results for two simulated electron numbers within the given error bars. For completeness, we note that we find average signs of $S=0.0587(3)$ and $S=0.3405(3)$ for $N=34$ and $N=14$, which explains the larger level of statistical uncertainty in the former case. Next, we consider the low-density low-temperature case of $r_s=50$ and $\Theta=1$ in Fig.~\ref{fig:FiniteSize2}. These conditions correspond to the strongest coupling parameter considered in this work and, thus, serve well to study the impact of the system size on the correlation structure. For these results, we found it most useful to separate the datasets for different values of $\tau$ and the top, center and bottom panels of Fig.~\ref{fig:FiniteSize2} corresponding to $\tau=0$, $\tau=\beta/10$ and $\tau=\beta/2$, respectively. First, the cancellation of positive and negative terms due to the fermionic anti-symmetry under particle coordinate exchange plays a very minor role is this regime (see Sec.~\ref{sec:sign_problem} for a more elaborate discussion), and the statistical uncertainty is very low for both $N=34$ (green) and $N=14$ (red). Second, we find a negligible N-dependence, except for a single $\mathbf{q}$-point in the direct vicinity of the main peak. To ensure the convergence of the $N=34$ dataset, we have carried out an additional PIMC calculation for $N=70$ (yellow symbols). The results are in excellent agreement with the green points for all three depicted values of $\tau$.

We thus conclude that PIMC simulations with $N=34$ electrons are fully appropriate to study the structural and (imaginary-time) spectral properties of the 2DEG over the entire considered range of densities and temperatures.

\begin{figure}
    \centering
    \includegraphics[width=0.495\textwidth]{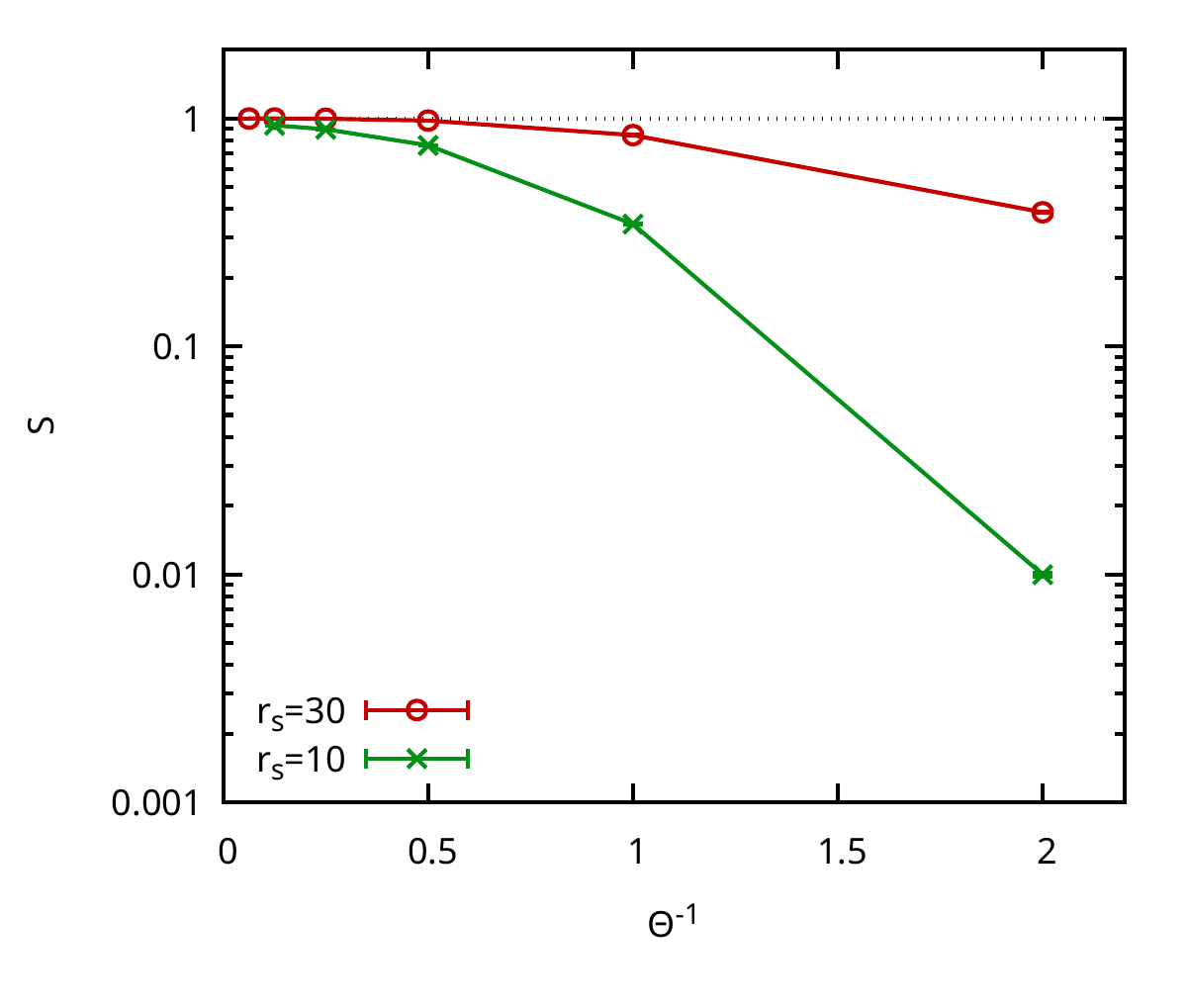}\\\includegraphics[width=0.495\textwidth]{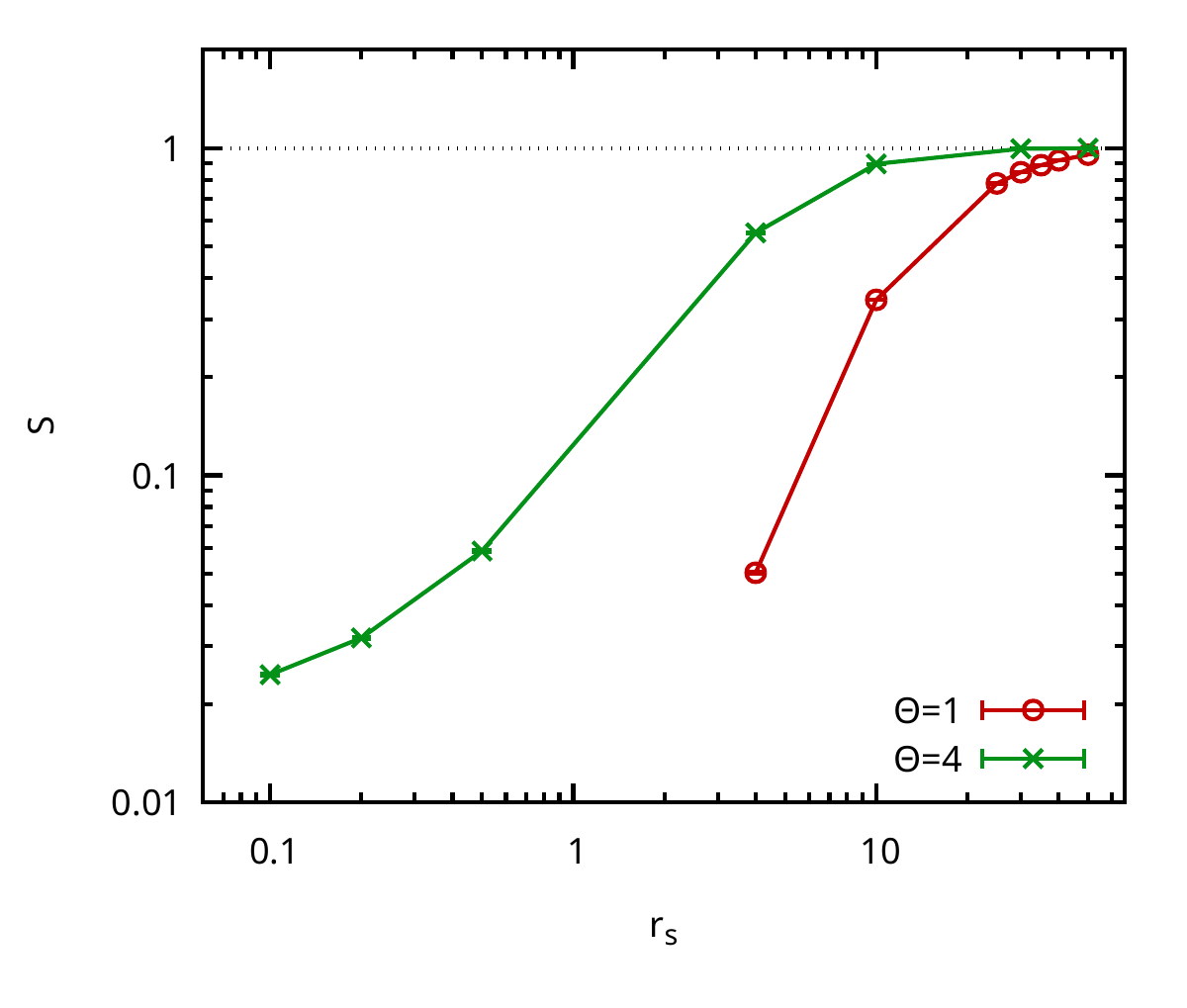}
    \caption{The average sign in PIMC simulations of the 2DEG with $N=34$ electrons. Top: dependence on the (inverse) temperature for $r_s=30$ (red circles) and $r_s=10$ (green crosses). Bottom: dependence on the density parameter $r_s$ for $\Theta=1$ (red circles) and $\Theta=4$ (green crosses).}
    \label{fig:sign}
\end{figure}

\subsection{Fermion sign problem\label{sec:sign_problem}}

Let us next briefly touch upon the fermion sign problem, which constitutes the central limiting factor of our simulations, in particular with respect to the temperature (lower bound) and the system size (upper bound). In the top panel of Fig.~\ref{fig:sign}, we show the dependence of $S$ on the (inverse) reduced temperature with $N=34$ electrons at $r_s=30$ (red circles) and $r_s=10$ (green crosses). In the high-temperature limit of $\Theta^{-1}\to0$, the sampled paths start to resemble classical point particles and the frequency of permutation cycles---the only source of sign changes in the configuration weight $W(\mathbf{X})$---within the simulation is low. As a consequence, both curves attain the expected limit of $S\to1$. Second, we find a substantial decrease of $S$ with decreasing temperature as the overlap of paths leads to a dramatic increase in pair exchanges and, hence, sign changes. Third, we find that the sign problem is substantially less severe for $r_s=30$ compared to $r_s=10$ for equal values of $\Theta$. For the ideal case, the level of quantum degeneracy and, hence, the sign problem would be fully determined by $\Theta$ and, thus, independent of $r_s$. For the interacting UEG, on the other hand, the comparably stronger Coulomb repulsion at $r_s=30$ effectively separates individual electrons and, in this way, suppresses the formation of permutation cycles~\cite{Dornheim_permutation_cycles}.

This effect can be discerned even more clearly and systematically in the bottom panel of Fig.~\ref{fig:sign}, where we show the dependence of the average sign on $r_s$ for $\Theta=1$ (red circles) and $\Theta=4$ (green crosses), again for $N=34$ electrons.
In both cases, we find that $S$ attains its unity limit for low density and strong coupling, as expected. Conversely, we observe a sharp drop of $S$ upon increasing the density (decreasing $r_s$). This drop is mainly located around the intermediate coupling regime of $r_s\sim1-10$ where quantum degeneracy and Coulomb coupling effects are in direct competition with each other. In the high density limit of $r_s\to0$, the average sign converges to that of the ideal system at the same $\Theta$ that is low but finite. At the lower temperature of $\Theta=1$, simulations with $N=34$ electrons are only feasible for $r_s\gtrsim4$, for which we find an average sign of $S=0.0503(3)$.

\begin{figure}
    \centering
\includegraphics[width=0.495\textwidth]{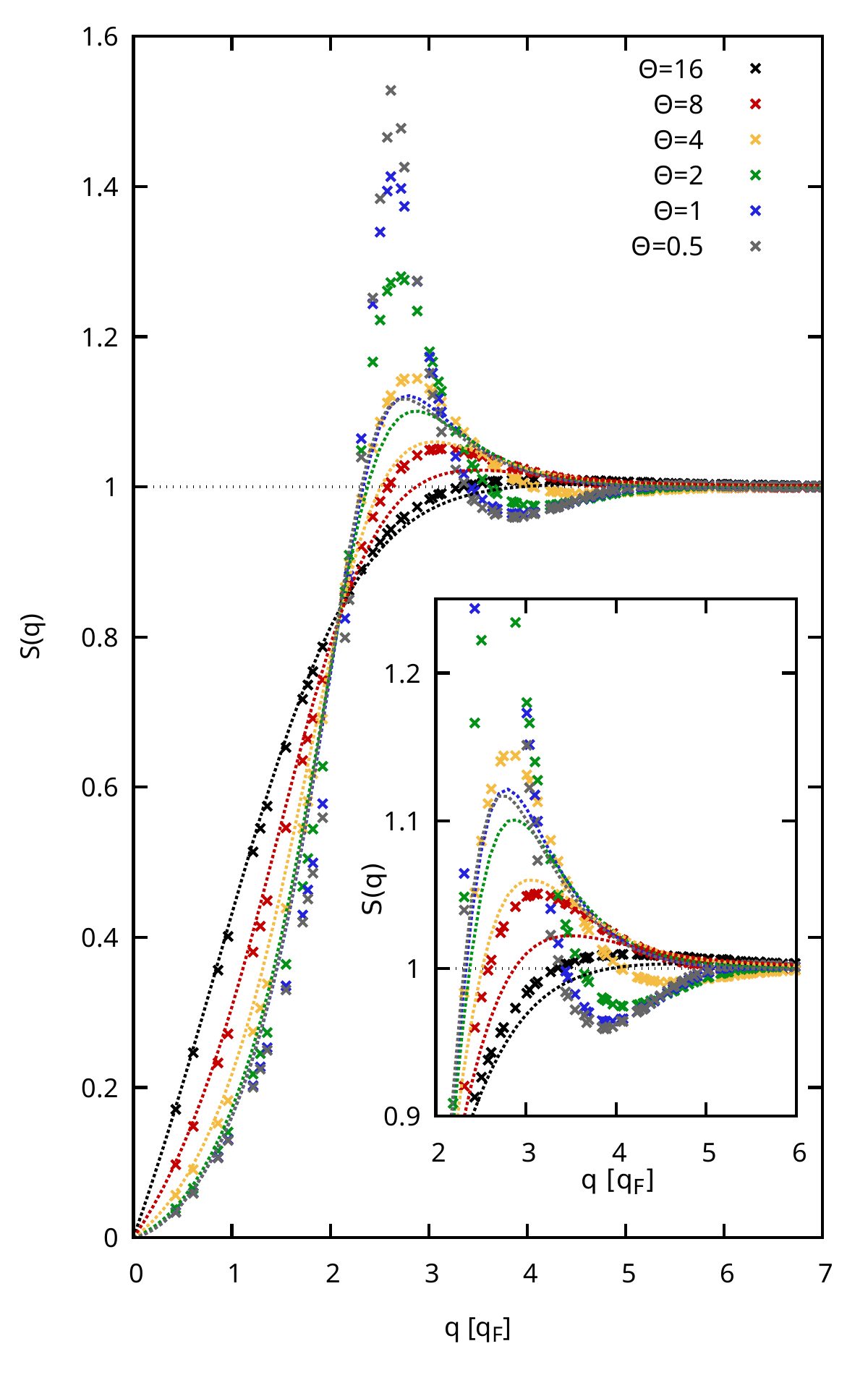} 
    \caption{The static structure factor $S(\mathbf{q})$ at $r_s=30$ and different reduced temperatures $\Theta$. Crosses: \emph{ab initio} PIMC simulations with $N=34$; dotted lines: 2D-HNC dielectric scheme \cite{kalkavouras2026dielectricformalism2duniform}. The inset depicts a magnified segment.    }
    \label{fig:structure_rs30}
\end{figure}

\begin{figure*}
    \centering
\includegraphics[width=0.495\textwidth]{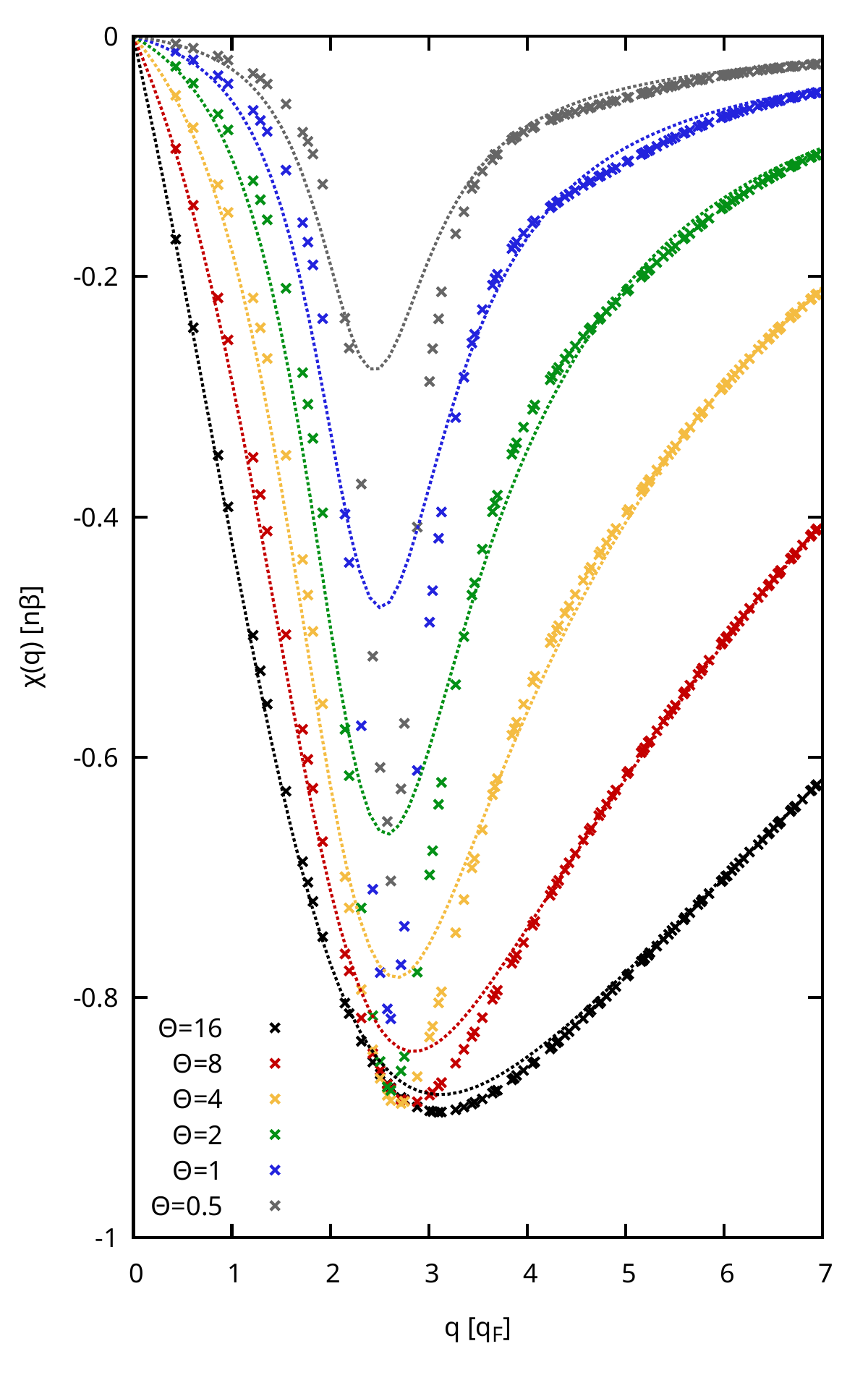} 
\includegraphics[width=0.495\textwidth]{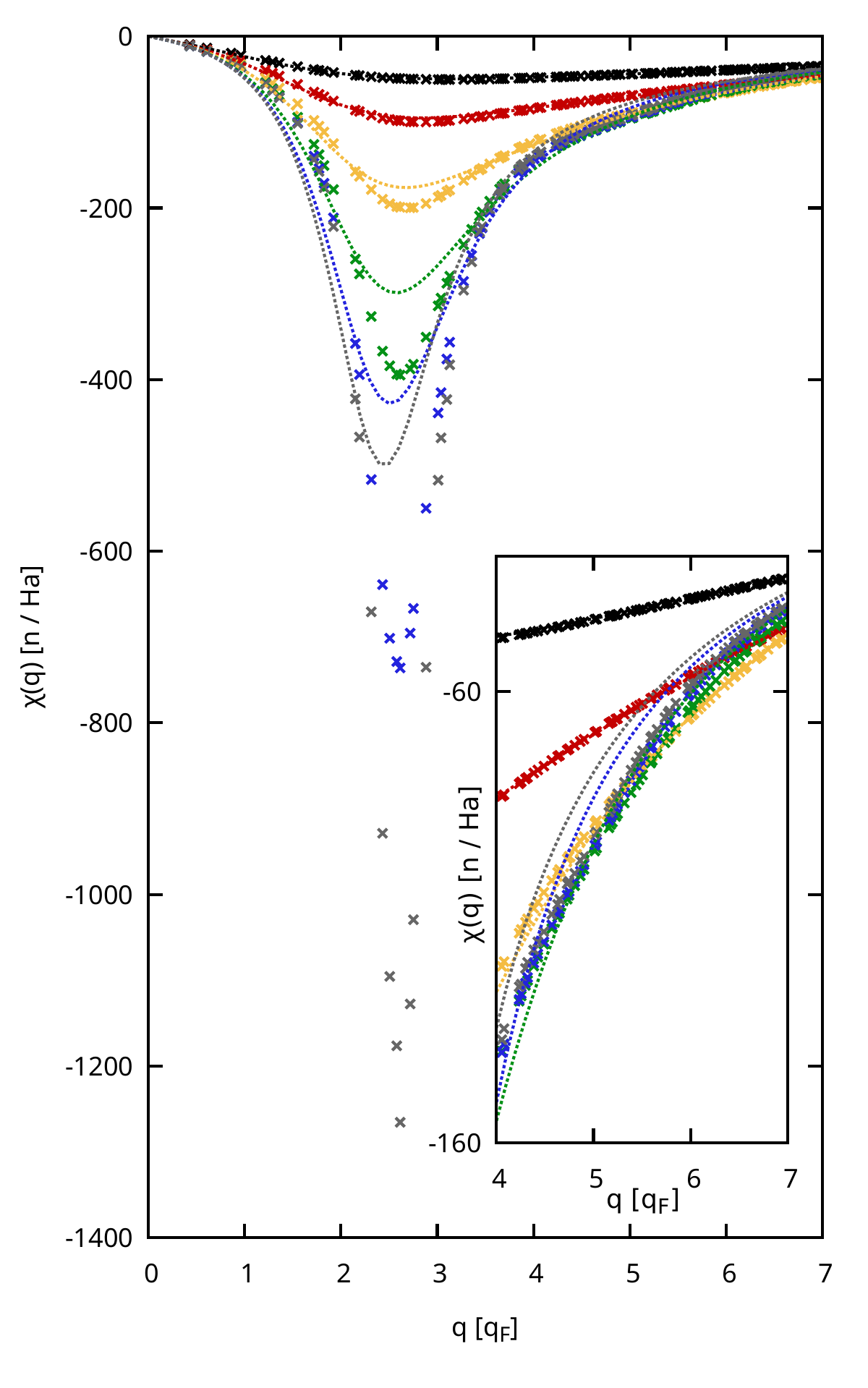} 
    \caption{The static linear density response function $\chi(\mathbf{q})$ at $r_s=30$ and different values of the reduced temperature $\Theta$. Crosses: \emph{ab initio} PIMC simulations with $N=34$, computed via Eq.~(\ref{eq:static_chi}); dotted lines: 2D-HNC dielectric scheme~\cite{kalkavouras2026dielectricformalism2duniform}. Left: $\chi(\mathbf{q})$ normalized with respect to the inverse temperature. Right: $\chi(\mathbf{q})$ unnormalized with respect to the inverse temperature. The inset in the right panel depicts a magnified segment around the short wavelength asymptote.    }
    \label{fig:response_rs30}
\end{figure*}

\subsection{Structural properties and density response\label{sec:structure}}

Let us next briefly touch upon the structural and linear density response properties of the 2DEG. Since an extensive parameter scan of the static structure factor $S(\mathbf{q})$ and static density response $\chi(\mathbf{q})$ has already been presented by Kalkavouras \emph{et al.}~\cite{kalkavouras2026dielectricformalism2duniform}, we here restrict ourselves to a temperature scan in the strongly coupled electron liquid regime at $r_s=30$.
Fig.~\ref{fig:structure_rs30} shows $S(\mathbf{q})$ for six different values of $\Theta$ in the range of $\Theta=0.5,\dots,16$; the crosses show our new PIMC results and the dotted lines the nominally most accurate dielectric theory that currently exists for the 2DEG, namely the hypernetted chain (HNC) closure for the static local field correction. 

From a physics perspective, the PIMC results nicely cover the transition from a structureless gas at $\Theta=16$ (black) to a strongly coupled liquid at $\Theta=0.5$ (grey). Qualitatively, this transition happens around $\Theta=2$ for this density, where the first peak exceeds the often quoted value of $S(\mathbf{q})=1.2$~\cite{Kundu_POP_2014}. In addition, we find a significant minimum in the vicinity of $q\approx4q_\textnormal{F}$ for $\Theta\gtrsim4$, but, interestingly, the magnitude of the second peak does not exceed $S(\mathbf{q})=1.003$ even for $\Theta=0.5$.

\begin{figure*}
    \centering
 \includegraphics[width=0.45\textwidth]{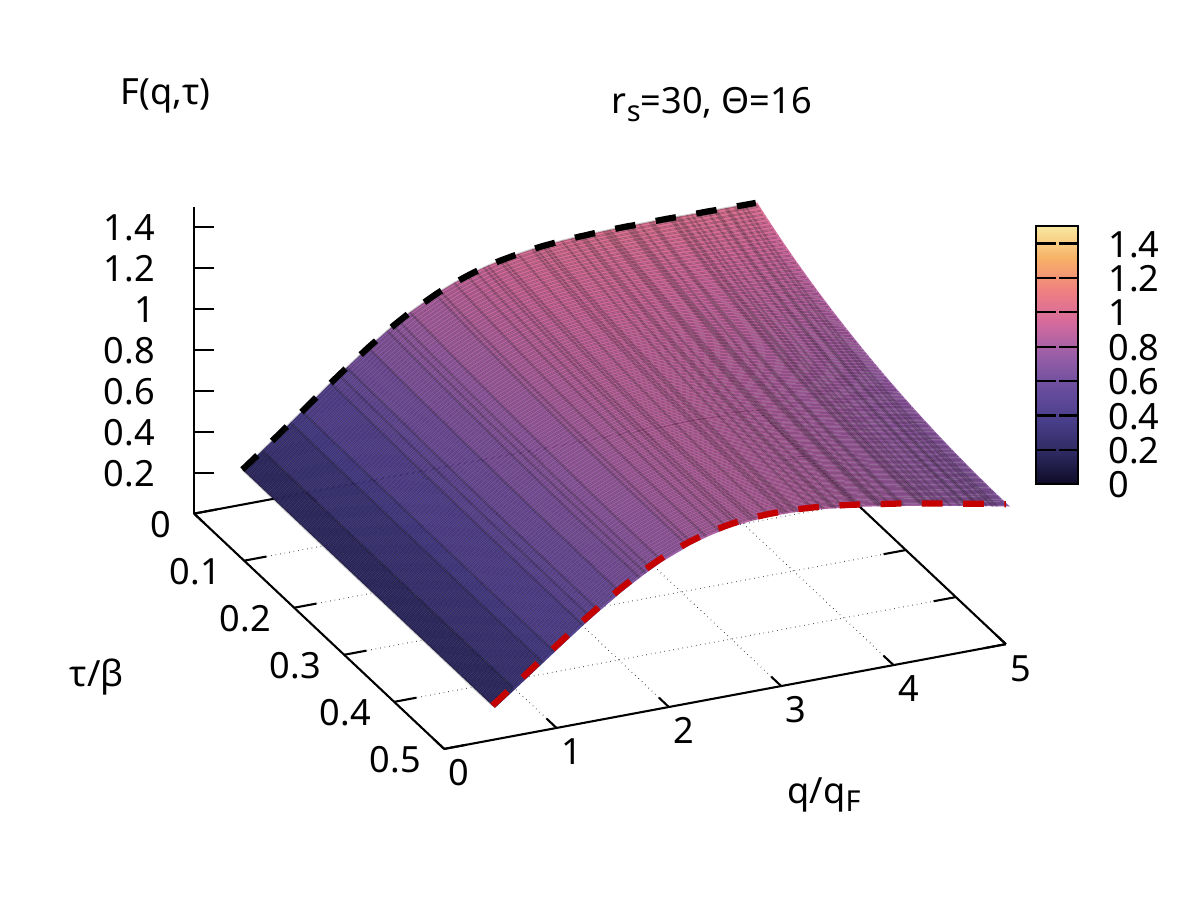}\includegraphics[width=0.45\textwidth]{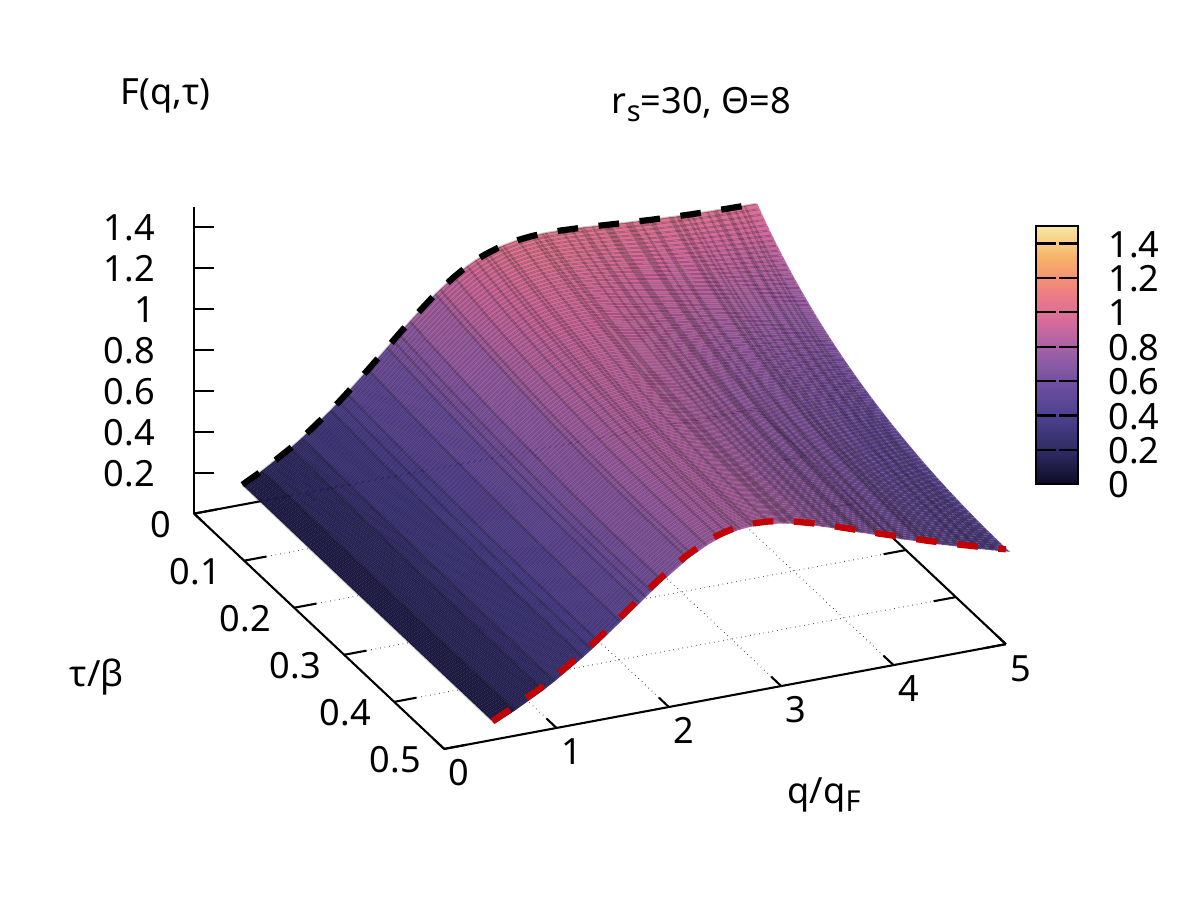}\\\vspace*{-1cm}\includegraphics[width=0.45\textwidth]{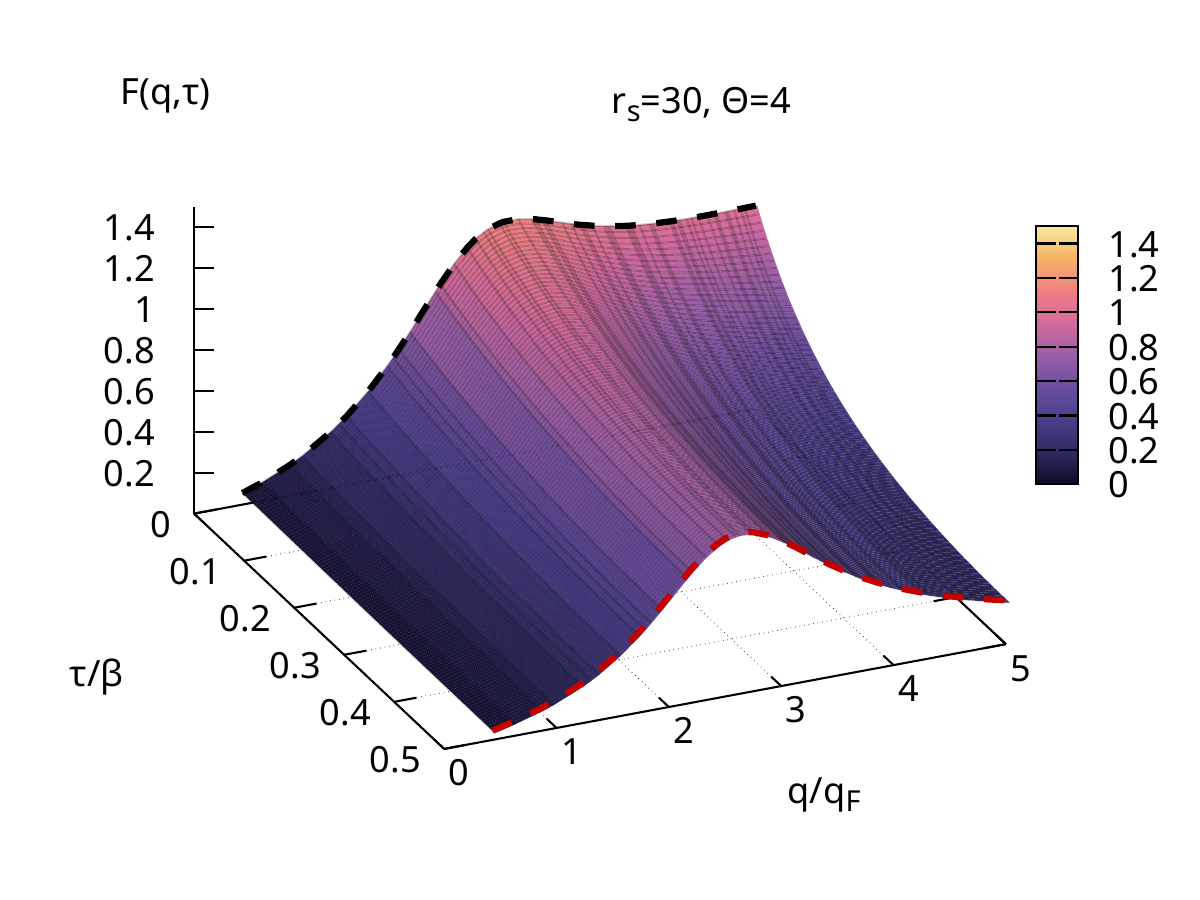}\includegraphics[width=0.45\textwidth]{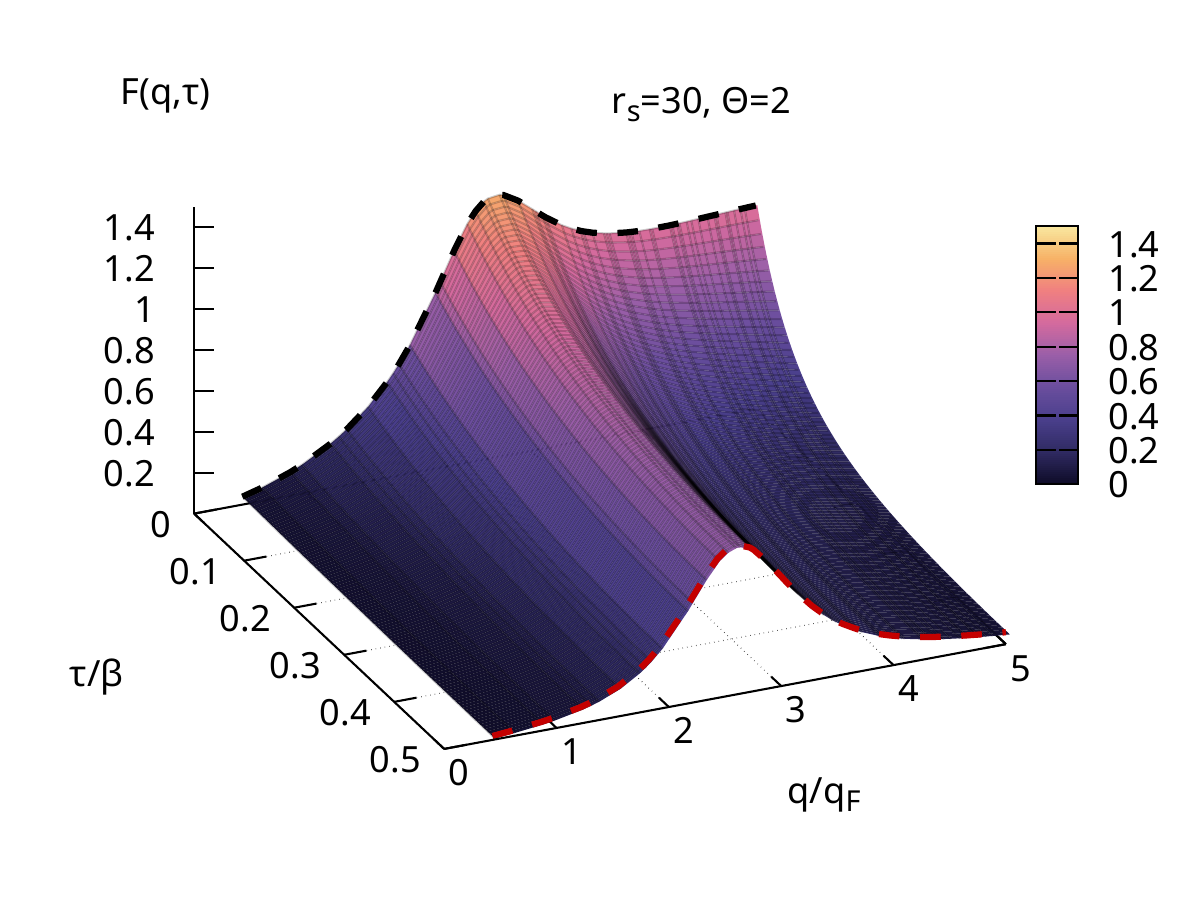}\\\vspace*{-1cm}\includegraphics[width=0.45\textwidth]{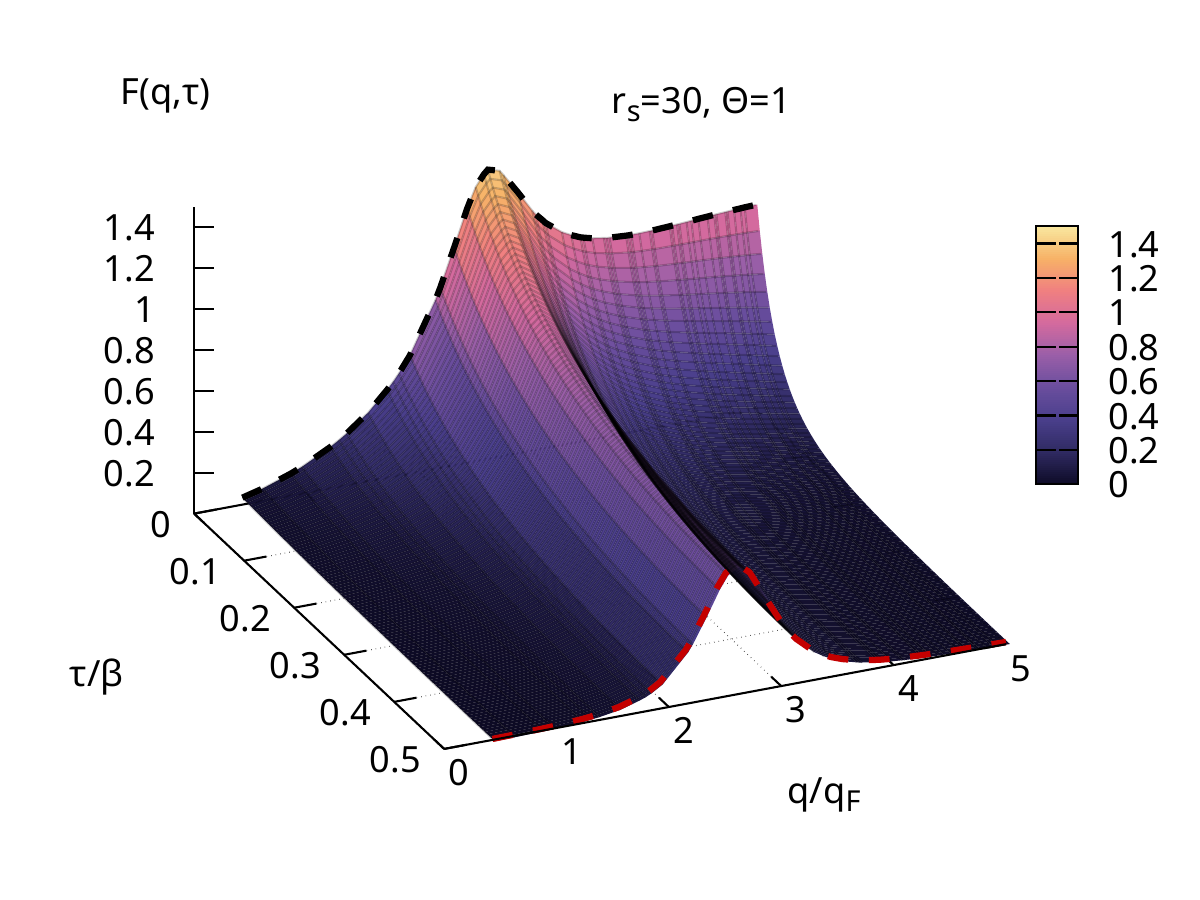}\includegraphics[width=0.45\textwidth]{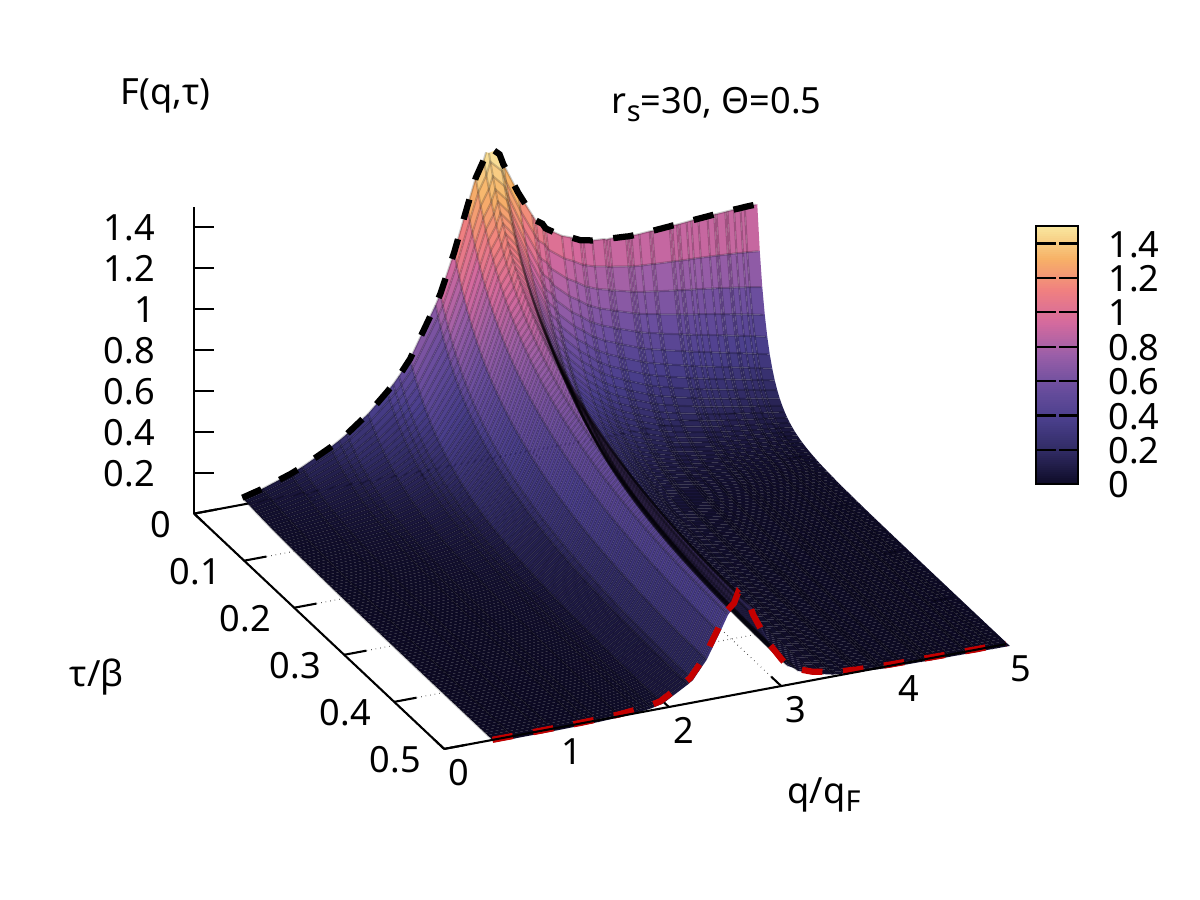}
    \caption{\emph{Ab initio} PIMC results for the ITCF $F(\mathbf{q},\tau)$ of the 2DEG at $r_s=30$ and a broad range of temperatures. The dashed black and solid red lines indicate the static structure factor $S(\mathbf{q})=F(\mathbf{q},0)$ and the thermal structure factor $S_{\beta/2}(\mathbf{q})=F(\mathbf{q},\beta/2)$, respectively. 
    }
    \label{fig:3D_rs30}
\end{figure*}

\begin{figure*}
    \centering
\includegraphics[width=0.45\textwidth]{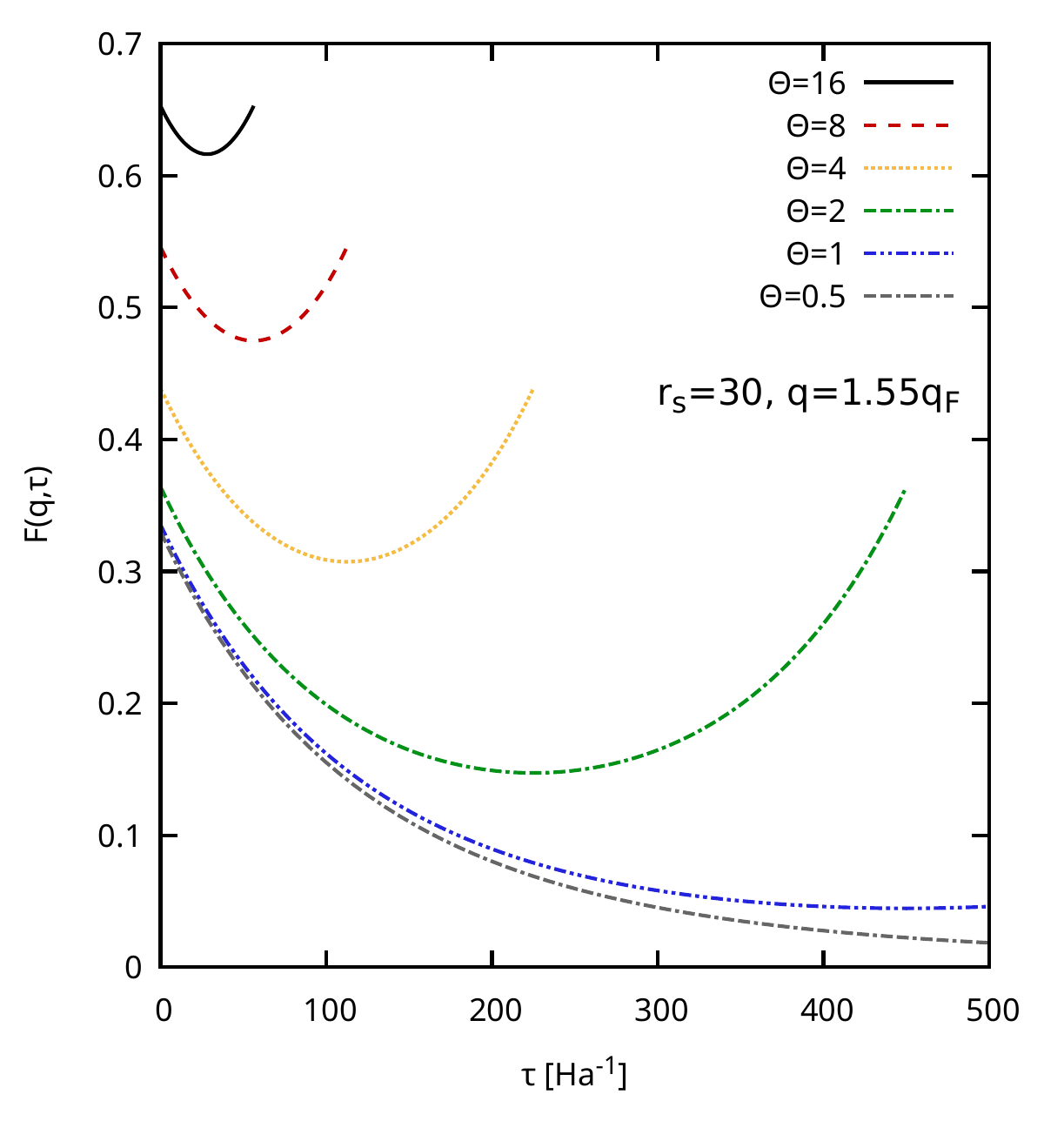}\includegraphics[width=0.45\textwidth]{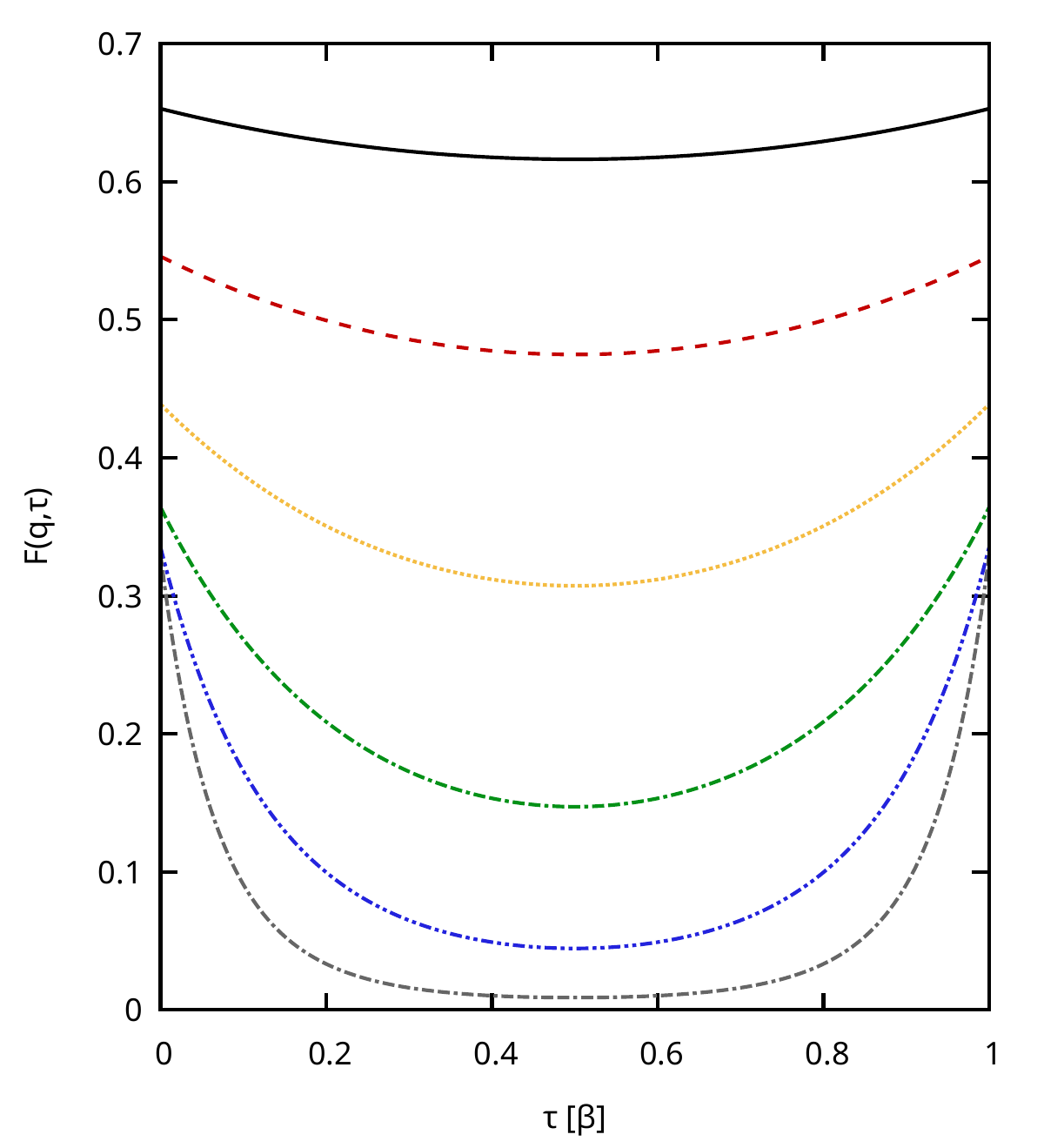}  
    \caption{ \emph{Ab initio} PIMC results for the ITCF $F(\mathbf{q},\tau)$ of the 2DEG at $r_s=30$ for $q=1.55q_\textnormal{F}$, with the colors and line styles distinguishing different values of the reduced temperature $\Theta$. Left: imaginary-time axis in absolute units; right: imaginary-time axis normalized to the inverse temperature $\beta$.    }
    \label{fig:ITCF_rs30_curves}
\end{figure*}

The dielectric HNC results capture the correct short wavelength limit $S(q\to\infty)=1$ as well as the exact long wavelength asymptote of~\cite{kalkavouras2026dielectricformalism2duniform}
\begin{eqnarray}\label{eq:SSF0}
    S(q\to0) = \frac{q}{2\pi n \beta}\ .
\end{eqnarray}
In the intermediate wavenumber regime that is shaped by exchange--correlation effects, we find significant differences towards the PIMC reference results even at the highest considered temperature, where $S(\mathbf{q})$ is relatively featureless. At lower temperatures (and, hence, stronger coupling), the quality of the HNC results noticeably deteriorates. Overall, we find that, while the position of the first peak of $S(\mathbf{q})$ is captured relatively accurately, its height is dramatically underestimated. In addition, the static HNC closure completely misses the first minimum.

In Fig.~\ref{fig:response_rs30}, we repeat this analysis for the static linear density response function $\chi(\mathbf{q})$, and the left panel has been normalized with respect to the inverse temperature $\beta$. In this way, the depth of the pronounced minimum, that appears around $q\approx2.5-3q_\textnormal{F}$, is broadly comparable between temperatures in our PIMC results. Interestingly, the HNC does not reproduce this trend, and underestimates the depth of the minimum by a factor of three at the lowest temperature, whereas the underestimation is only slight at $\Theta=16$. To understand this observation, we recall the physical origin of the minimum and of its position, which can be illustrated from two different, yet consistent perspectives.
First, we note that $\chi(\mathbf{q})$ decays in the limit of $q\to0$ because of screening~\cite{kalkavouras2026dielectricformalism2duniform},
\begin{eqnarray}\label{eq:chi0}
    \chi(q\to0) = - \frac{q}{2\pi}\ ,
\end{eqnarray}
and also decays in the limit of $q\gg q_\textnormal{F}$ because of quantum delocalization effects. The latter can be readily seen by re-calling the Matsubara representation of the static structure factor~\cite{stls,Tolias_JCP_2024,Dornheim_PRB_2024},
\begin{eqnarray}\label{eq:Matsubara}
    S(\mathbf{q}) = - \frac{1}{n\beta} \sum_{l=-\infty}^\infty \widetilde{\chi}(\mathbf{q},z_l)\ ,
\end{eqnarray}
with $z_l=i2\pi l \beta^{-1}$ the bosonic Matsubara frequencies. In the classical limit, there are no $l\neq0$ terms, giving the simple correspondence between $S(\mathbf{q})$ and $\chi(\mathbf{q})=\widetilde{\chi}(\mathbf{q},0)$:
\begin{eqnarray}
    \chi_\textnormal{cl}(\mathbf{q}) = - n \beta S_\textnormal{cl}(\mathbf{q})\ ;
\end{eqnarray}
since $S_\textnormal{cl}(q\to\infty)=1$, it is trivial that $\chi_\textnormal{cl}(q\to\infty)=-n\beta$. For quantum systems, on the other hand, terms with $l\neq0$ start to play an increasingly important role with increasing $q$, shifting weight from $\chi(\mathbf{q})$ to other Matsubara frequencies, whereas the total sum in Eq.~(\ref{eq:Matsubara}) must still attain unity for $q\gg q_\textnormal{F}$.
The quantum delocalization origin of the decay of $\chi(\mathbf{q})$ for large $q$ also immediately becomes apparent when we re-call the imaginary-time version of the fluctuation--dissipation theorem, Eq.~(\ref{eq:static_chi}), which has been used to compute all PIMC results that are shown in Fig.~\ref{fig:response_rs30}.
Specifically, Eq.~(\ref{eq:static_chi}) states that $\chi(\mathbf{q})$ is directly proportional to the area under the ITCF $F(\mathbf{q},\tau)$; the latter decays increasingly fast with $\tau$ for large $q$ (see, e.g., Fig.~\ref{fig:3D_rs30} and Sec.\ref{sec:spectral} more broadly), thus also explaining the observed trend.
Coming back to the position of the minimum of $\chi(\mathbf{q})$ at intermediate wavenumbers, we can now view it as a sweat spot away from the decay of $\chi(\mathbf{q})$ in the limits of large and small $q$. More to the point, Dornheim \emph{et al.}~\cite{Dornheim_Nature_2022}
have explained the $\chi(\mathbf{q})$ minimum with a pair alignment model, which posits that the system reacts most strongly when $\lambda=2\pi/q$ is commensurate with a spatial pattern that minimizes the interaction energy in the system. The minimum in $\chi(\mathbf{q})$ thus broadly coincides when $\lambda$ is comparable to the interparticle distance $d$, which is indeed what we find. It is important to note that this pair alignment is an exchange--correlation effect and, as such, particularly difficult to capture with approximate dielectric theories such as the static HNC.

Returning to results shown in the left panel of Fig.~\ref{fig:response_rs30}, we note significant differences between the PIMC data and HNC curves around $q=5q_\textnormal{F}$ for $\Theta\lesssim2$, i.e., in the vicinity of twice the wavenumber of the minimum of $\chi(\mathbf{q})$. This discrepancy was absent in the static structure factor $S(\mathbf{q})$ and must reflect a spectral feature encoded into the full $\tau$-dependence of the ITCF $F(\mathbf{q},\tau)$. In fact, a comparably larger absolute magnitude of $\chi(\mathbf{q})$ without a corresponding maximum in $S(\mathbf{q})=F(\mathbf{q},0)$ indicates a reduced decay with respect to $\tau$ of $F(\mathbf{q},\tau)$ and, considering Eq.~(\ref{eq:Laplace}) above, a shift of spectral weight to lower frequencies by necessity. Indeed, Chuna \emph{et al.}~\cite{Chuna_JCP_2025} have recently investigated this feature in the 3D UEG and attributed it to a second roton feature that indicates the incipient formation of a phonon dispersion. We will explore this possibility in more detail in Sec.\ref{sec:spectral}.

Finally, the right panel of Fig.~\ref{fig:response_rs30} shows the same information, but in physical units, i.e., without normalizing with respect to $\beta$. In this representation, it becomes clear that the physical system will display the same response to the external density perturbation in both the long and short wavelength limits; see also the inset showing a magnified segment around convergence to the latter. We also nicely observe the strong increase in the magnitude of the density response at intermediate wavenumbers upon decreasing the temperature due to the increased importance of exchange--correlation effects.

\begin{figure}
    \centering
\includegraphics[width=0.45\textwidth]{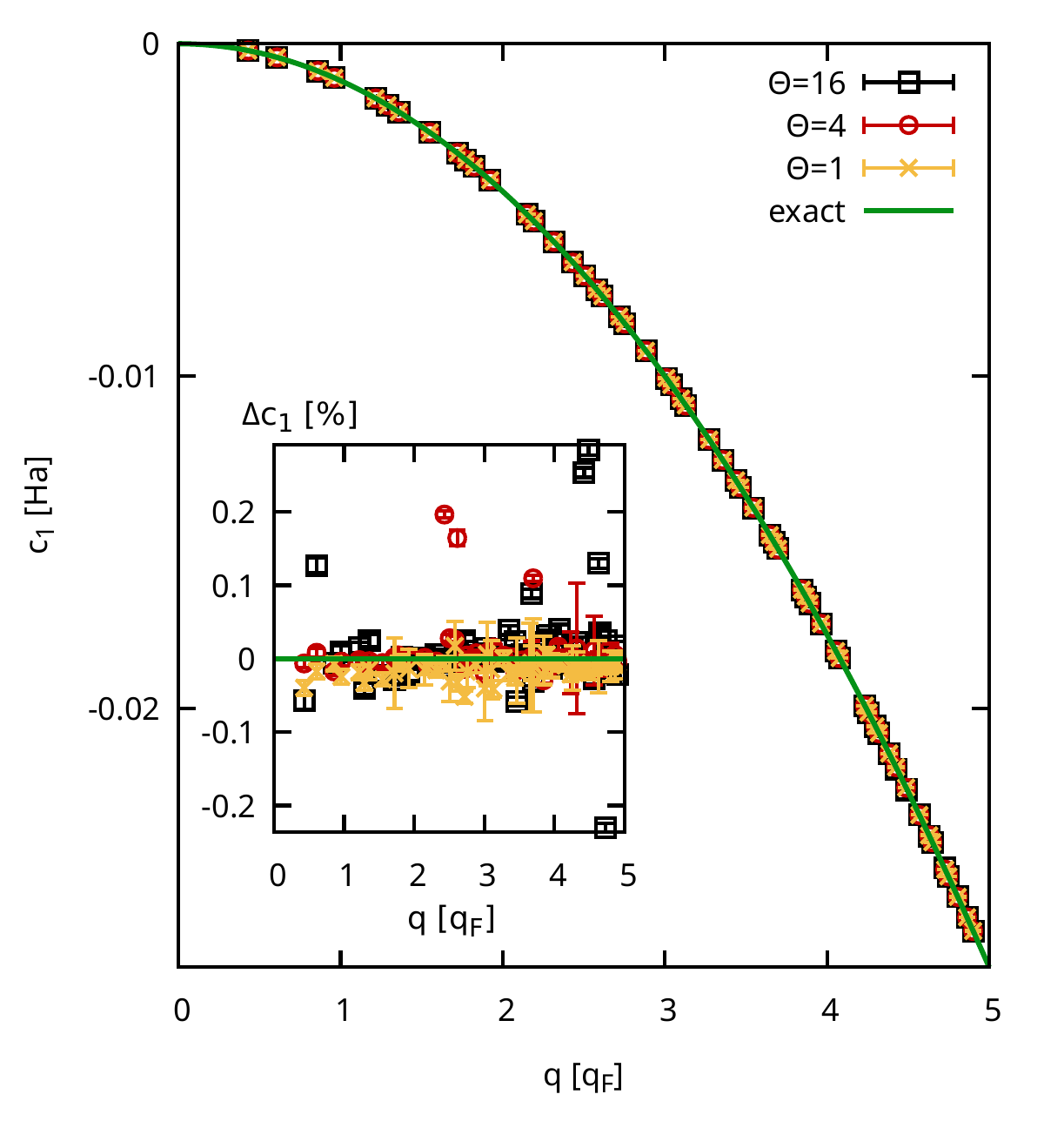} 
    \caption{The first frequency moment of the dynamic structure factor [Eq.~(\ref{eq:define_moments})] for the 2DEG at $r_s=30$ and different values of the reduced temperature $\Theta$. The black squares, red circles and yellow crosses have been obtained from polynomial fits to the ITCF, cf.~Eq.~(\ref{eq:Taylor}). Solid green: the exact f-sum rule. The inset shows the relative deviation between the extracted PIMC results and the exact curve in percent.
    }
    \label{fig:fsum}
\end{figure}

\begin{figure}
    \centering
\includegraphics[width=0.45\textwidth]{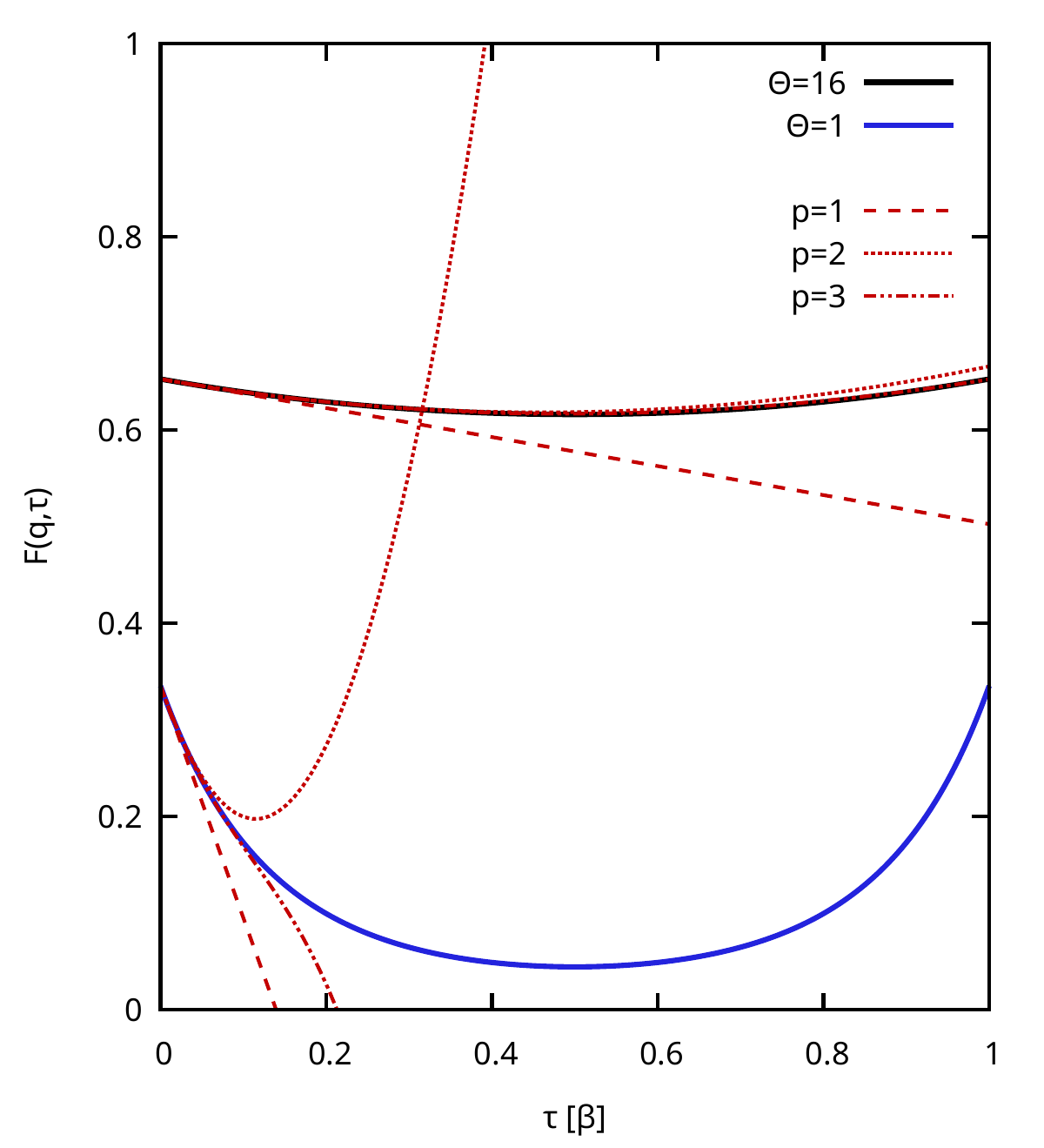} 
    \caption{Results for the ITCF $F(\mathbf{q},\tau)$ of the 2DEG at $r_s=30$ and $\Theta=16$ (black), $\Theta=1$ (blue) for $q=1.55q_\textnormal{F}$. Solid lines: \emph{ab initio} PIMC simulations. Dashed, dotted, dashed-double-dotted red lines: polynomial expansion [Eq.~(\ref{eq:Taylor})] evaluated up to the order $p=1$, $p=2$ and $p=3$, respectively.
    }
    \label{fig:coeff}
\end{figure}

\subsection{Imaginary time density--density correlation function\label{sec:spectral}}

To investigate the spectral properties of the 2DEG, we first consider the full $q$- and $\tau$-dependence of the ITCF, which is shown in Fig.~\ref{fig:3D_rs30} at $r_s=30$ and a broad range of temperatures $\Theta$. We note that it is sufficient to restrict ourselves to the half-plane of $\tau\in[0,\beta/2]$ due to the well-known symmetry of the ITCF,
\begin{eqnarray}\label{eq:ITCF_symmetry}
    F(\mathbf{q},\tau) = F(\mathbf{q},\beta-\tau)\ ,
\end{eqnarray}
which directly follows from the detailed balance of the dynamic structure factor, $S(\mathbf{q},-\omega)=S(\mathbf{q},\omega)e^{-\beta\omega}$ in thermal equilibrium~\cite{Dornheim_MRE_2023,Dornheim_T_2022,Dornheim_T_follow_up}. The dashed black curve corresponds to the static structure factor $S(\mathbf{q})=F(\mathbf{q},0)$, which was already discussed in Sec.~\ref{sec:structure} above. The basic question, thus, concerns the dependence of $F(\mathbf{q},\tau)$ on $\tau$, which, as we shall see, contains some subtleties. The main trend that is, at least to some degree, present in all panels of Fig.~\ref{fig:3D_rs30} is the increasingly fast decay of the ITCF for large $q$. This trend can be understood by re-calling the PIMC implementation of $F(\mathbf{q},\tau)$ as a correlation function of single-particle densities at different imaginary times [Eq.~(\ref{eq:define_ITCF})], see also Fig.~\ref{fig:scheme}. For large $q$ and thus small $\lambda$, increasingly small $\tau$-differences are sufficient to reduce correlations along the imaginary-time diffusion process.
In Fig.~\ref{fig:3D_rs30}, we followed the common convention and normalized the $\tau$-axis with respect to $\beta$, which gives one the impression of a faster $\tau$-decay for lower temperatures. However, this observation is somewhat deceptive. In the left and right panels of Fig.~\ref{fig:ITCF_rs30_curves}, we show the $\tau$-dependence of the ITCF for a fixed wavenumber of $q=1.55q_\textnormal{F}$ for the same conditions as in Fig.~\ref{fig:3D_rs30} with an absolute (unnormalized) $\tau$-axis and with the same normalized $\tau$-axis like before, respectively.

\begin{figure*}
    \centering
 \includegraphics[width=0.45\textwidth]{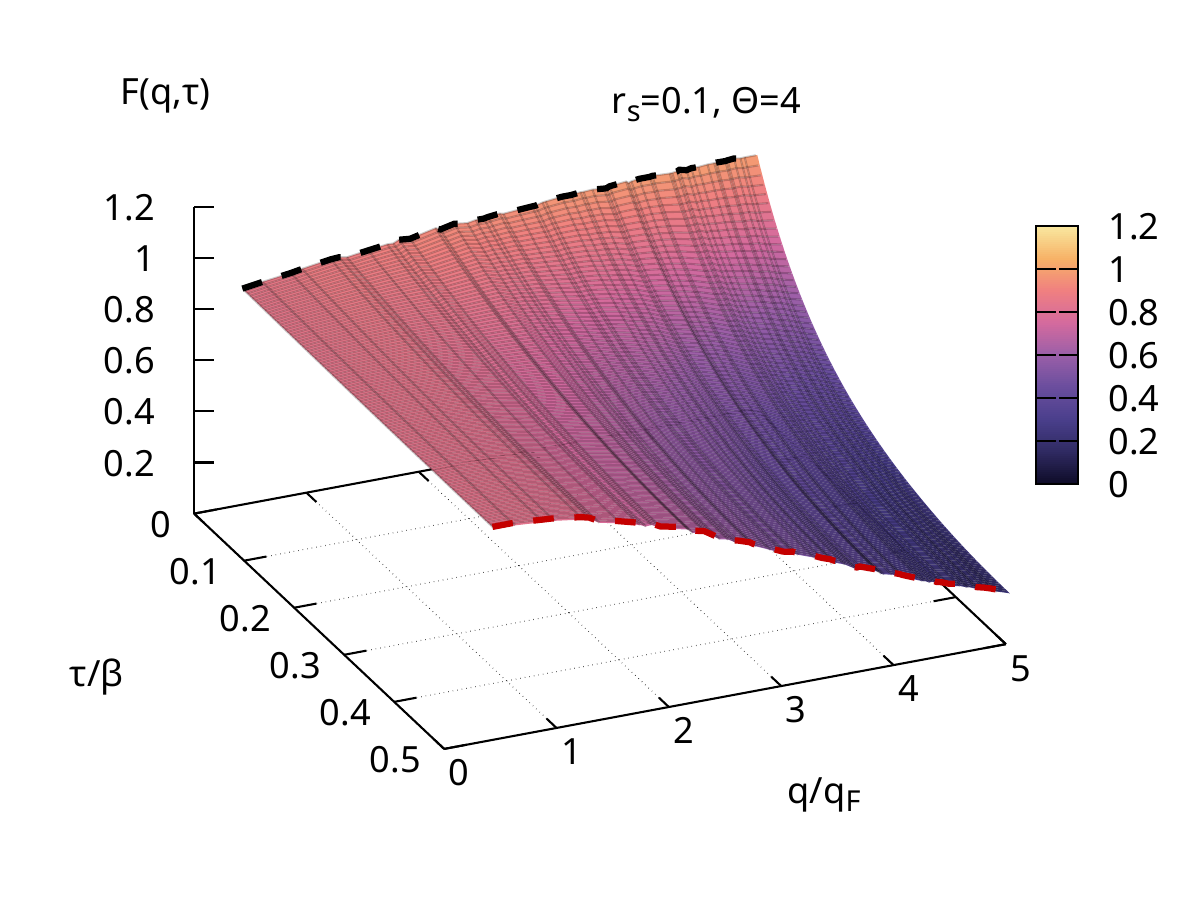}\includegraphics[width=0.45\textwidth]{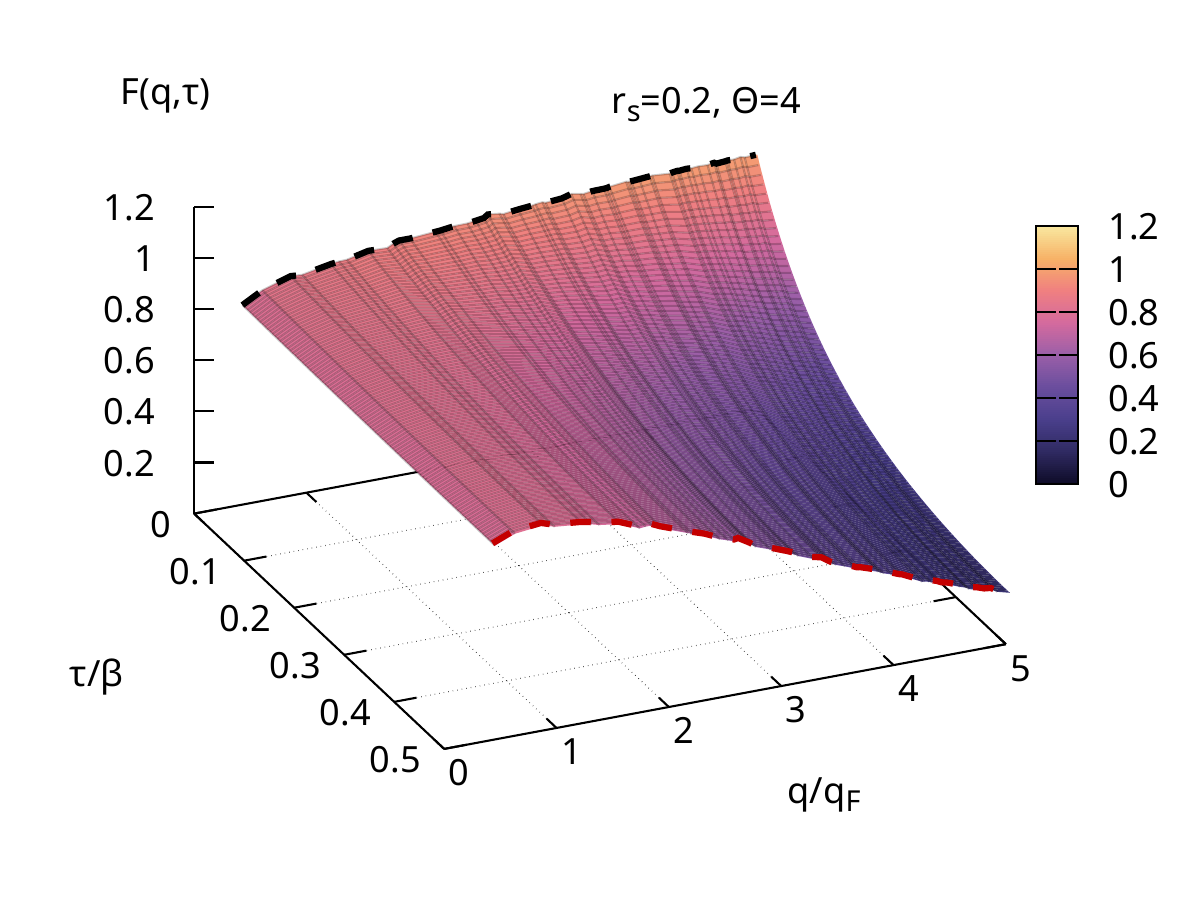}\\\vspace*{-1cm}\includegraphics[width=0.45\textwidth]{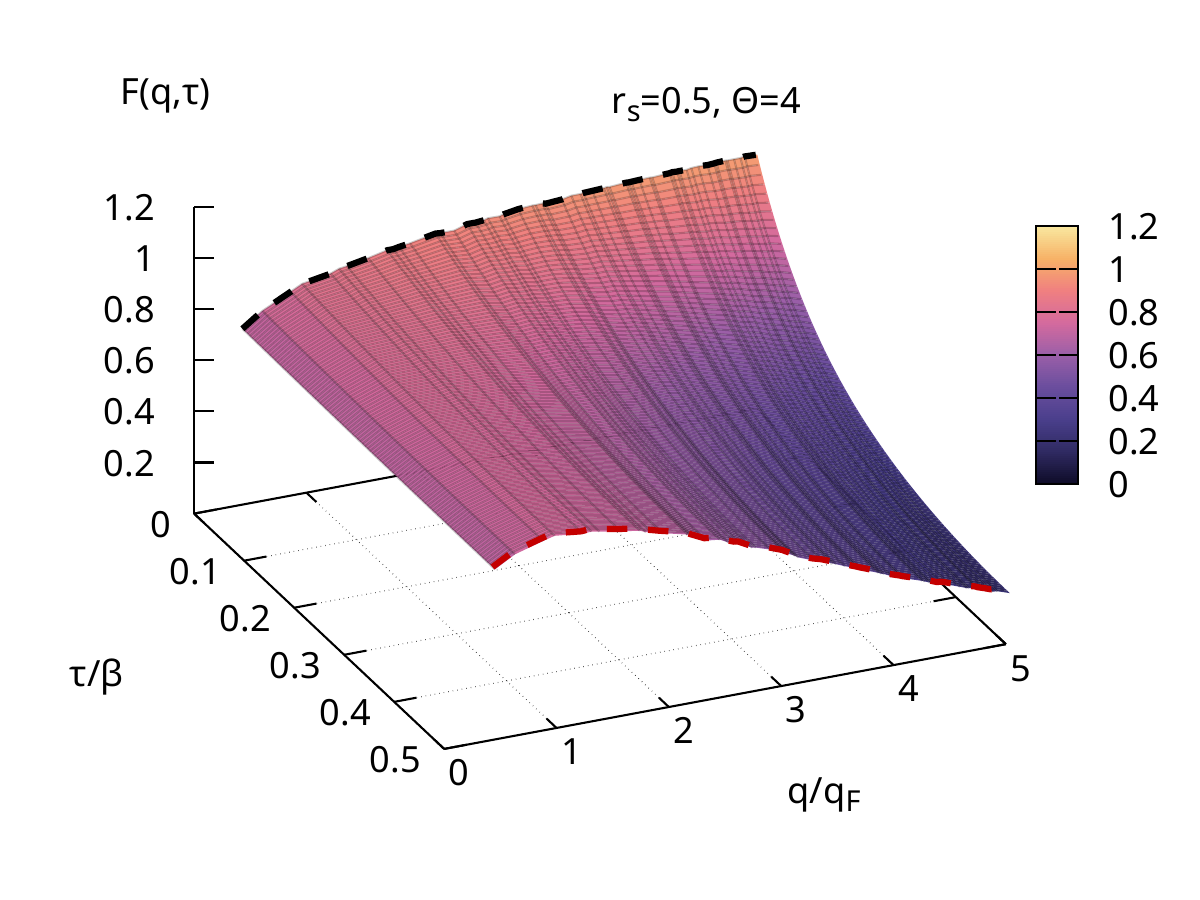}\includegraphics[width=0.45\textwidth]{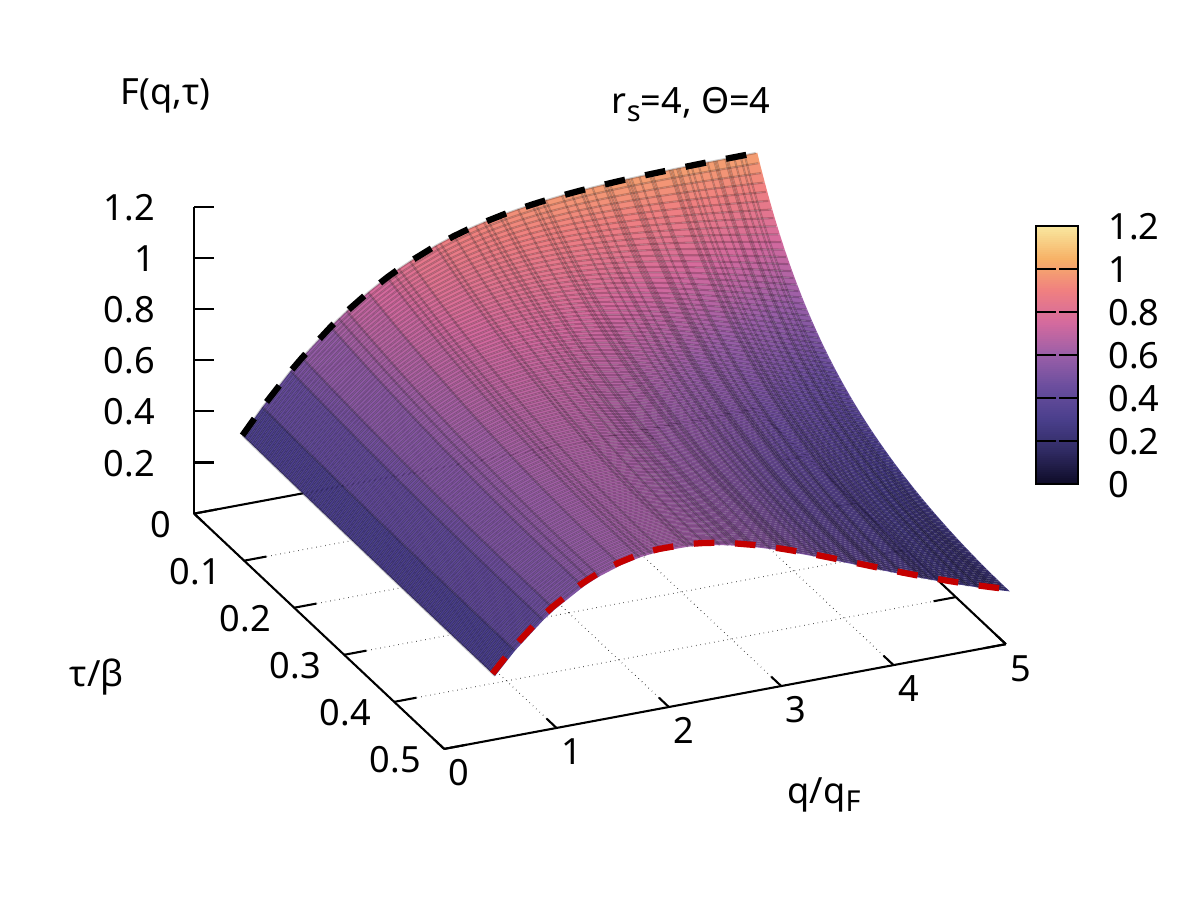}\\\vspace*{-1cm}\includegraphics[width=0.45\textwidth]{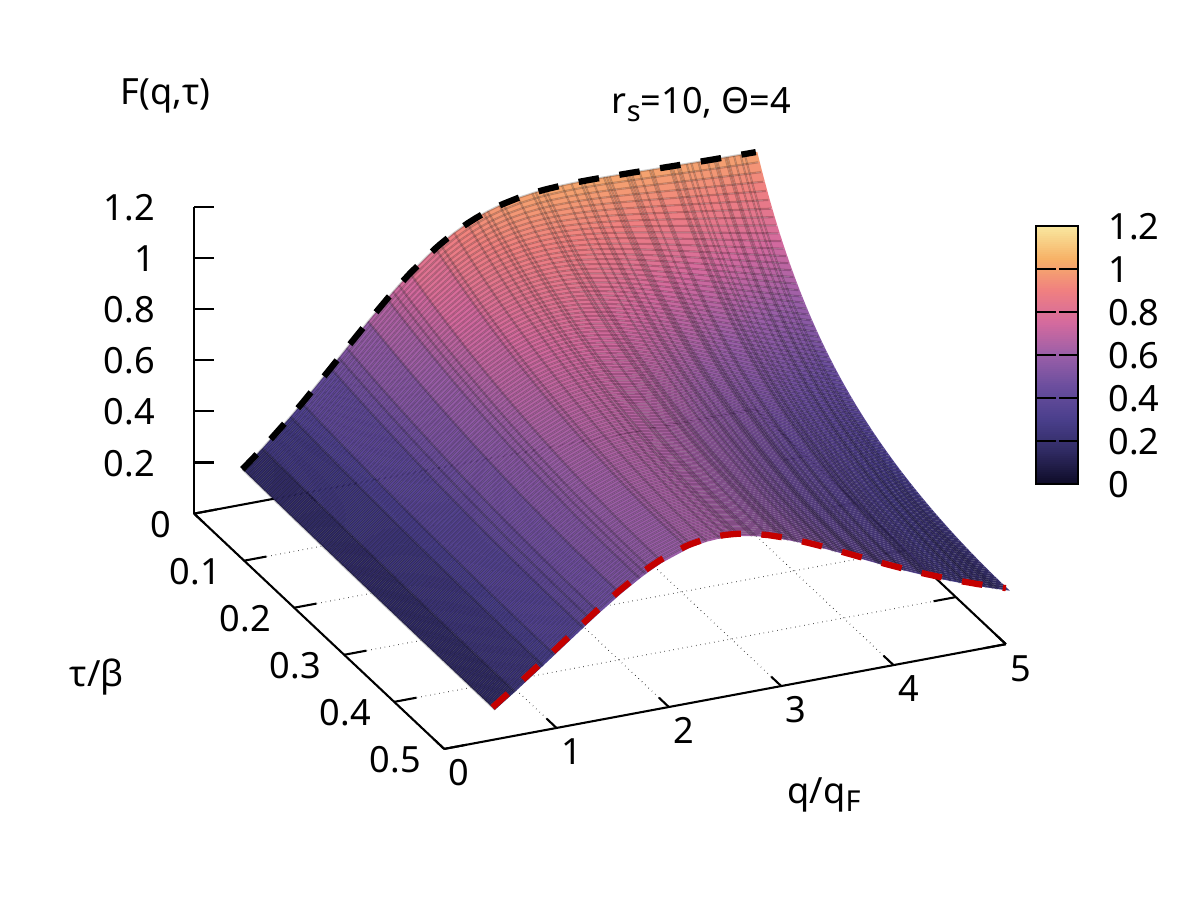}\includegraphics[width=0.45\textwidth]{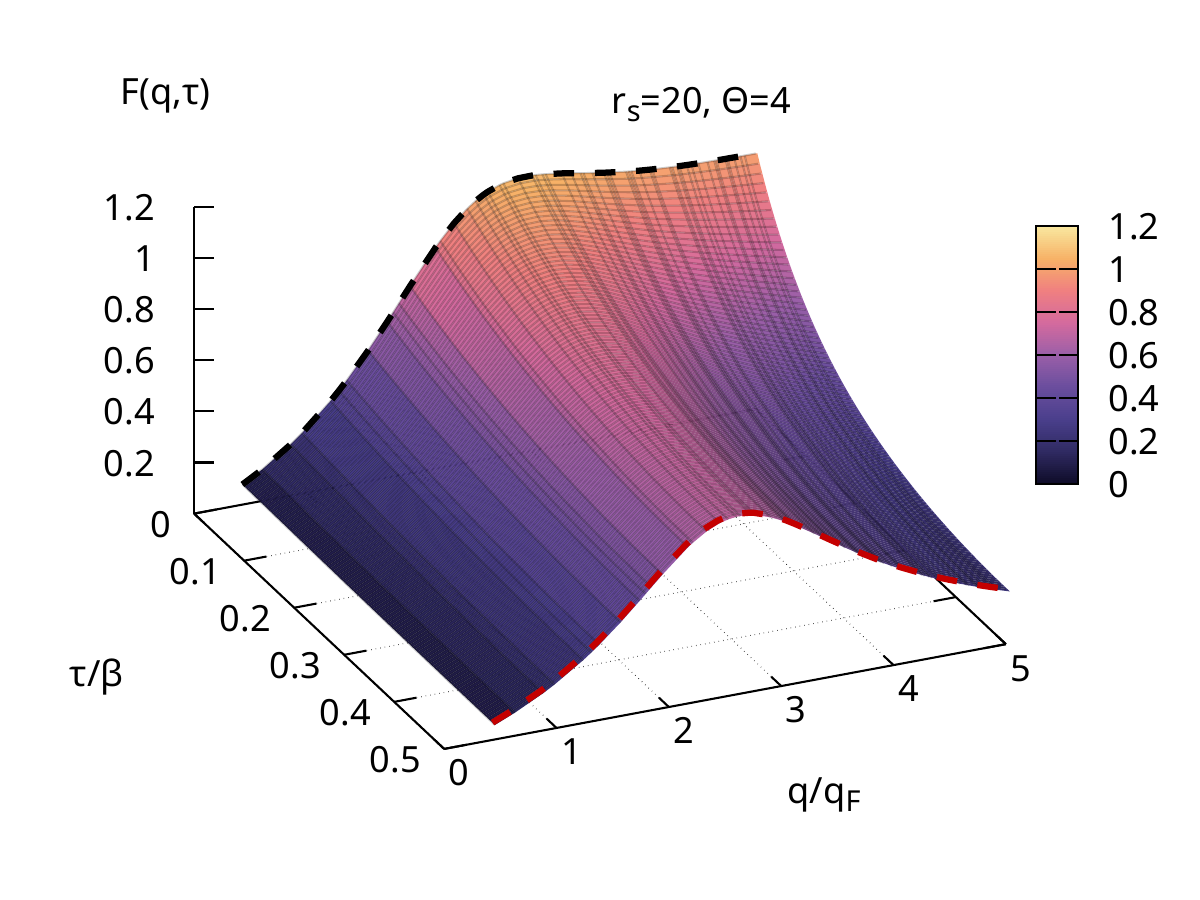}\\\vspace*{-1cm}\includegraphics[width=0.45\textwidth]{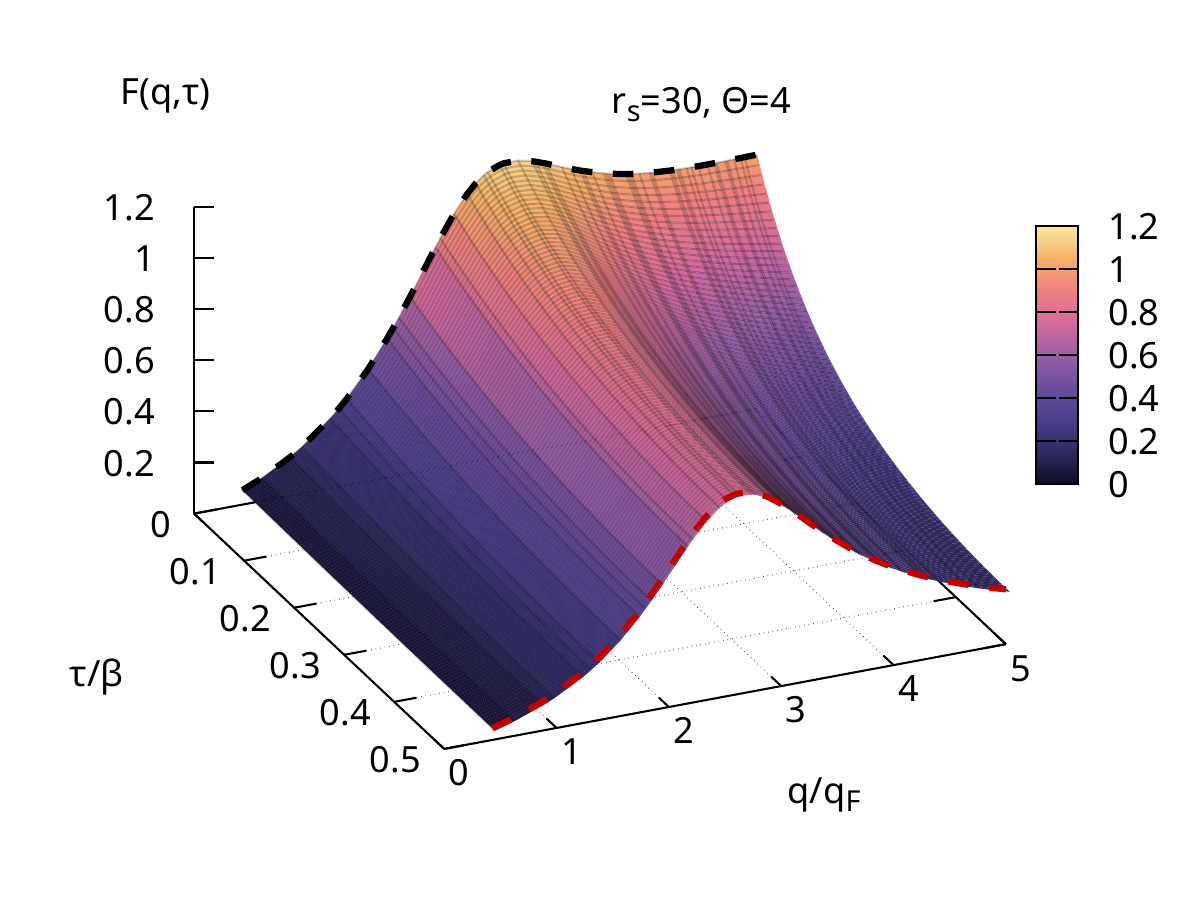}\includegraphics[width=0.45\textwidth]{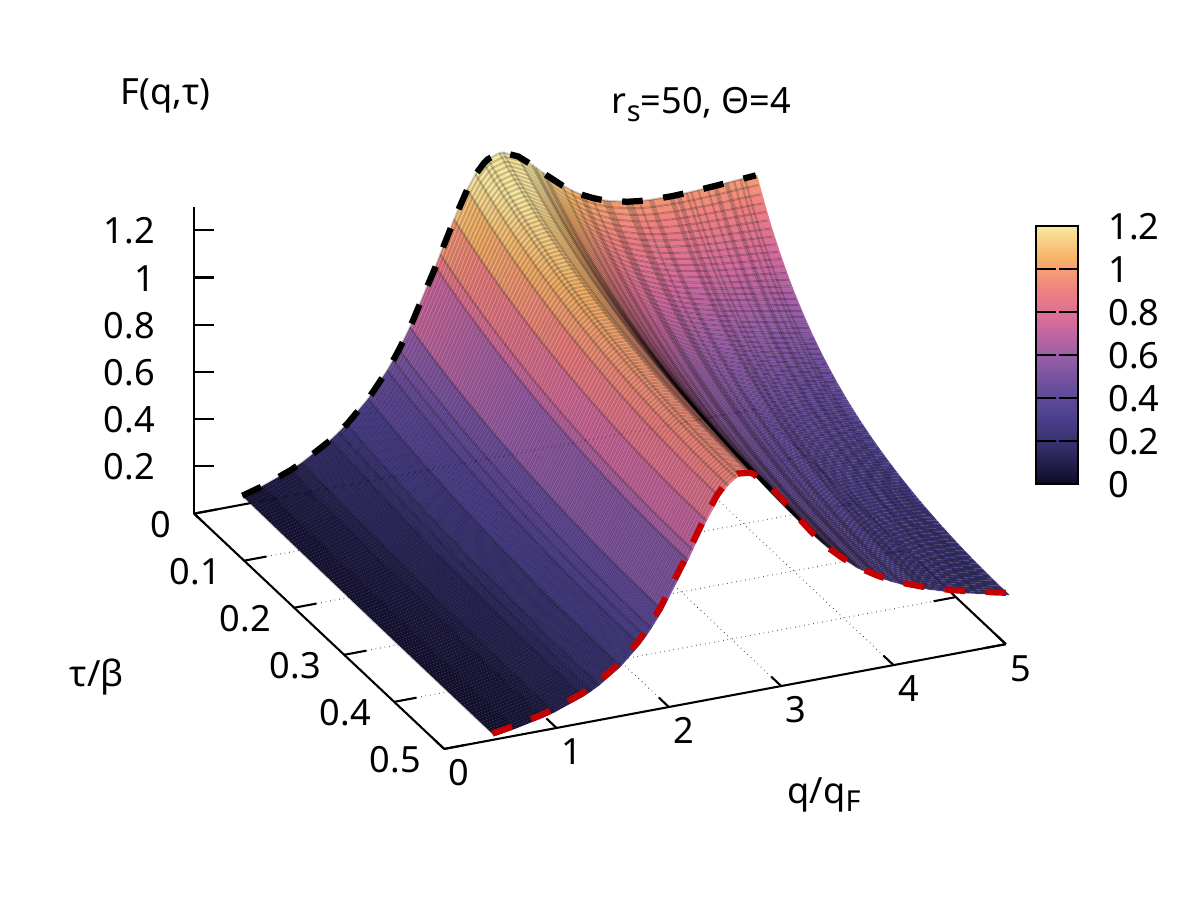}%\includegraphics[width=0.45\textwidth]{2SPLOT3D_2DUEG_N34_rs30_theta4_P200_NBIN100.pdf}
    \caption{\emph{Ab initio} PIMC results for the ITCF $F(\mathbf{q},\tau)$ at $\Theta=4$ across a broad range of density parameters $r_s$. The dashed black and red lines indicate the static structure factor $S(\mathbf{q})=F(\mathbf{q},0)$ and thermal structure factor $S_{\beta/2}(\mathbf{q})=F(\mathbf{q},\beta/2)$, respectively. 
    }
    \label{fig:3D_T}
\end{figure*}

First, independent of the representation, all curves are monotonically ordered with respect to the temperature. This is analogous to the equivalent ordering of the static structure factor at this particular wavenumber, cf.~Fig.~\ref{fig:structure_rs30}.
Second, the left panel plainly illustrates the shorter $\tau$-intervals at higher temperatures, which are the main reason for the apparent reduced $\tau$-decay in Fig.~\ref{fig:3D_rs30} at large $\Theta$. In fact, the ITCF decays, to the linear order in $\tau$, equally fast independent of the temperature. This becomes apparent upon recalling the polynomial representation of $F(\mathbf{q},\tau)$ in Eq.~(\ref{eq:Taylor}), since the linear coefficient is directly related to the first $\tau$-derivative of $F(\mathbf{q},\tau)$ at $\tau=0$ and thus to the first frequency moment of $S(\mathbf{q},\omega)$, cf.~Eq.~(\ref{eq:define_moments}). The latter is determined by the f-sum rule,
\begin{eqnarray}\label{eq:fsum}
    M_S^{(1)}(\mathbf{q}) = \frac{\mathbf{q}^2}{2}\ ,
\end{eqnarray}
that does not depend on the temperature. To verify our data against this exact relation, we have performed polynomial fits to the PIMC results for $F(\mathbf{q},\tau)$ based on the approach introduced in Ref.~\cite{Dornheim_moments_2023}. Fig.~\ref{fig:fsum} shows the $q$-dependence of the thus extracted first coefficient, i.e., the first derivative of the ITCF corresponding to $-M_S^{(1)}(\mathbf{q})$, for $\Theta=16$ (black squares), $\Theta=4$ (red circle) and $\Theta=1$ (yellow crosses); indeed, no dependence on the temperature can be discerned with the naked eye. The solid green curve shows the exact derivative determined from Eq.~(\ref{eq:fsum}) that is in excellent agreement with the extracted PIMC results. The inset shows the relative deviations of the PIMC results from Eq.~(\ref{eq:fsum}). The error bars have been obtained with a standard leave-one-out binning procedure~\cite{berg2004markov} from a set of independent PIMC runs for each temperature. Overall, these error bars capture most fluctuations in the data, except for a few outliers. These reflect a small potential bias, e.g., from the truncation of the Taylor series of Eq.~(\ref{eq:Taylor}) and a potential compensation between different polynomial orders. Nevertheless, we find satisfactory agreement between PIMC and the exact f-sum rule within $\lesssim0.1\%$.

Additional details on the extraction of the ITCF $\tau$-derivatives and thus of the dynamic structure factor frequency moments are provided in Fig.~\ref{fig:coeff}, where we show $F(\mathbf{q},\tau)$ for $q=1.55q_\textnormal{F}$ at $\Theta=16$ (solid black line) and $\Theta=1$ (solid blue line). The dashed, dotted and dashed-double-dotted red curves correspond to the polynomial representation of Eq.~(\ref{eq:Taylor}), evaluated up to the first, second and third order, respectively. At $\Theta=16$, the ITCF is relatively flat and a cubic polynomial accurately describes $F(\mathbf{q},\tau)$ over the entire $\tau$-range. In stark contrast, the ITCF exhibits a more pronounced and more complex decay at $\Theta=1$. Therefore, a cubic polynomial remains accurate only over $\lesssim10\%$ of the relevant $\tau$-interval, and a higher polynomial order is needed for a meaningful fit. 

For completeness, we note that the frequency moments are interesting in their own right and have received considerable attention, e.g., for the normalization of experimental data~\cite{GarciaSaiz2008,Dornheim_SciRep_2024,Dornheim_NatComm_2025,Dornheim_POP_2025,gawne2026modelfreeinterpretationxraythomson}, for the estimation of the kinetic energy~\cite{kalkavouras2026kineticenergycubicsum}, and in the context of the analytic continuation, in particular via the Hamburger problem~\cite{tkachenko_book,Tkachenko_CPP_2018,Filinov_PRB_2023,Vorberger_PRL_2012}.

We finish our discussion of the ITCF with a dedicated analysis of its dependence on the coupling strength. In Fig.~\ref{fig:3D_T}, we show our new \emph{ab initio} PIMC results for the full ITCF $F(\mathbf{q},\tau)$ at $\Theta=4$ across a wide range of density parameters, ranging from $r_s=0.1$ (top left) to $r_s=50$ (bottom right). In the weakly coupled high-density limit, $F(\mathbf{q},\tau)$ appears to be relatively featureless apart from the by now familiar $\tau$-decay. This is somewhat deceptive as the static structure factor $S(\mathbf{q})=F(\mathbf{q},0)$ does abide by the linear asymptote given by Eq.~(\ref{eq:SSF0}), but our simulation box is too small to capture this long wavelength limit. With an increasing $r_s$, the ITCF starts to exhibit a more pronounced structure due to electronic exchange--correlation effects, with a small peak forming in $S(\mathbf{q})$ at $r_s=20$ for this value of $\Theta$. At $r_s=50$, this peak exceeds the previously quoted value of $1.2$~\cite{Kundu_POP_2014}, marking the onset of the strongly coupled liquid regime.

\begin{figure*}
    \centering
 \includegraphics[width=0.45\textwidth]{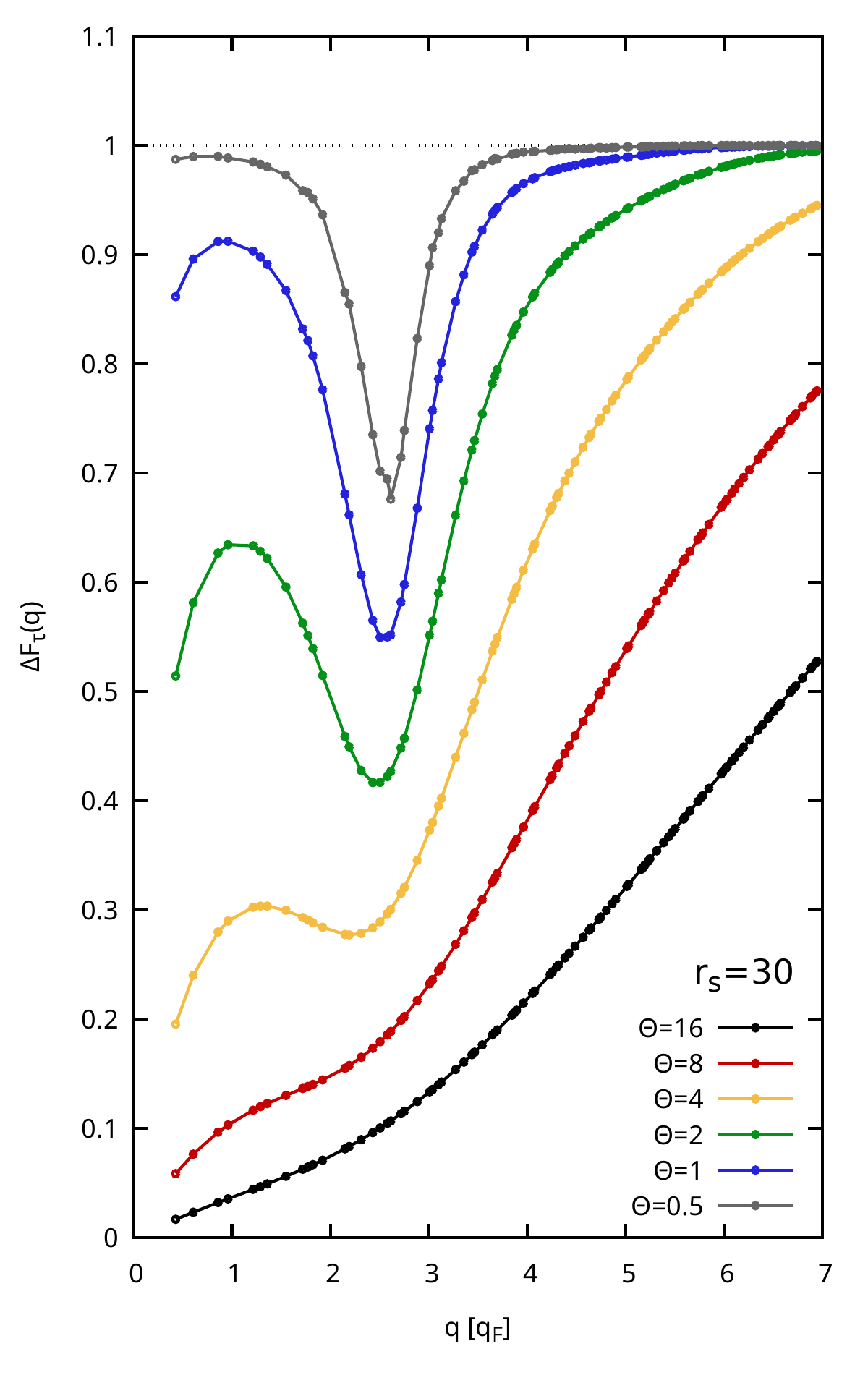}\includegraphics[width=0.45\textwidth]{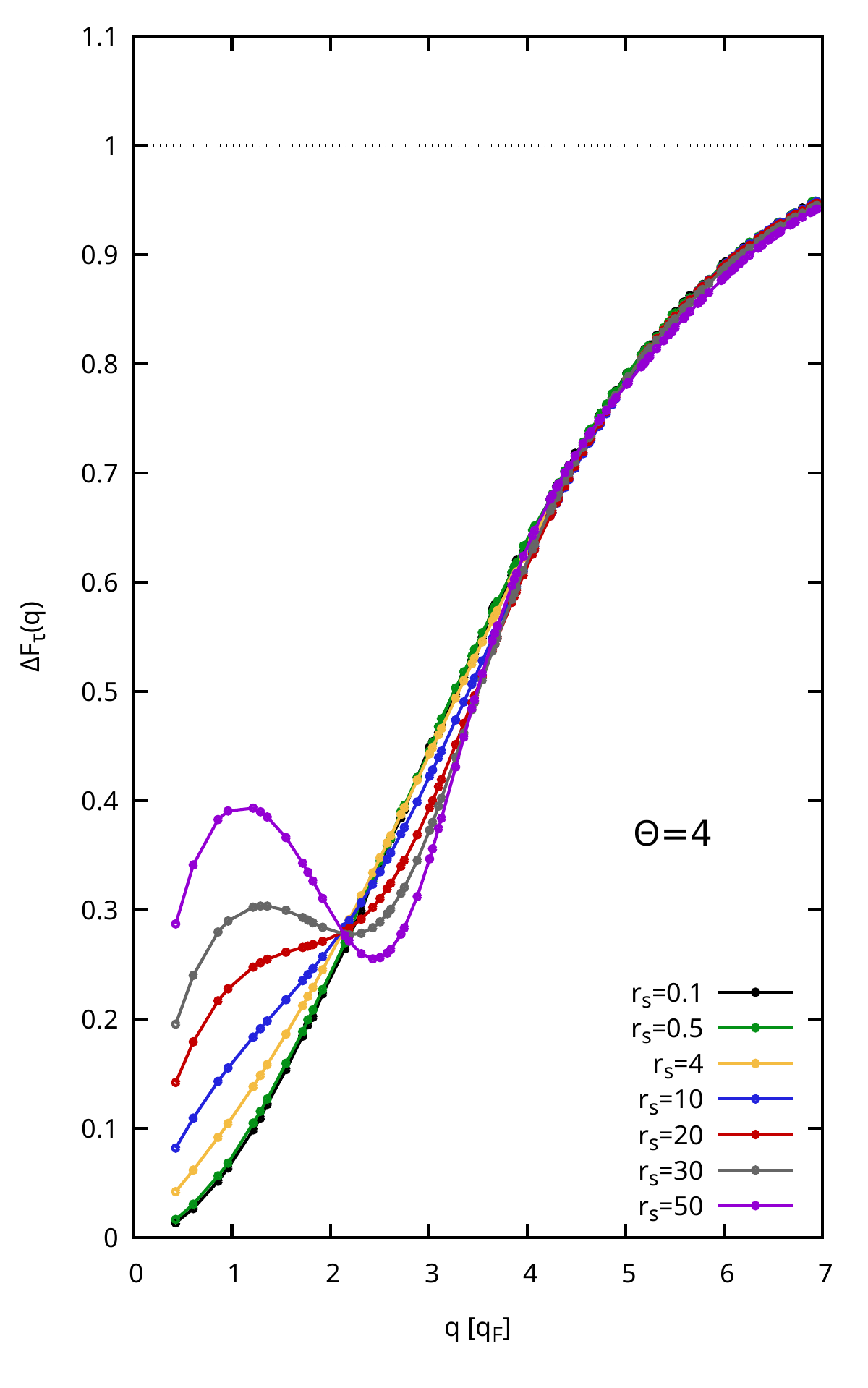}
    \caption{\emph{Ab initio} PIMC results for the ITCF $\tau$-decay measure $\Delta{F}_\tau(\mathbf{q})$ of the 2DEG divided by $S(\mathbf{q})=F(\mathbf{q},0)$ [Eq.~(\ref{eq:Delta_F_tau})] and evaluated for $\tau=\beta/2$. Left panel: $r_s=30$ for a range of different temperatures $\Theta=0.5,\dots,16$; right panel: $\Theta=4$ for a range of different density parameters $r_s=0.1,\dots,50$.}
    \label{fig:decay_rs30}
\end{figure*}

\begin{figure*}
    \centering
 \includegraphics[width=0.45\textwidth]{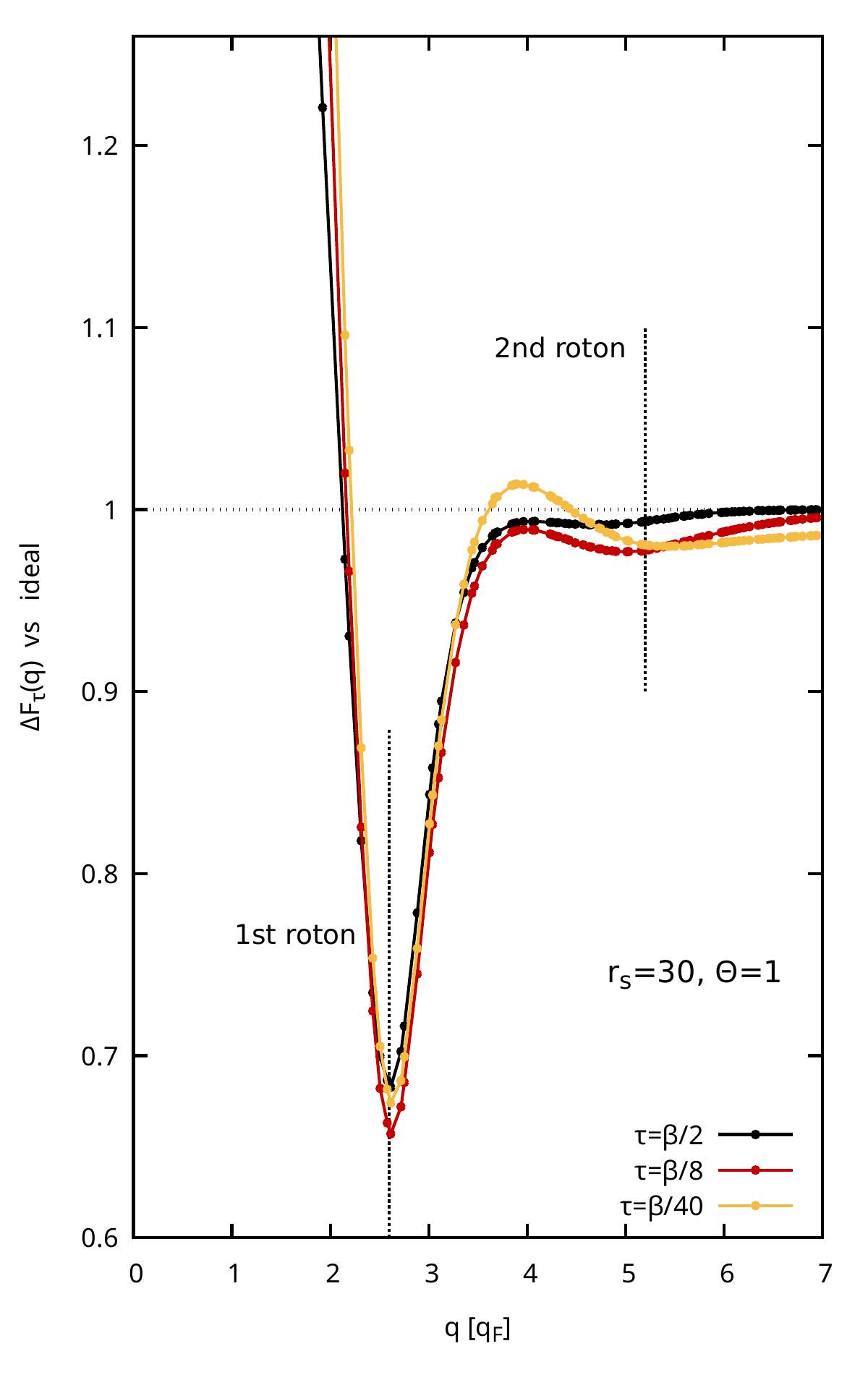}\includegraphics[width=0.45\textwidth]{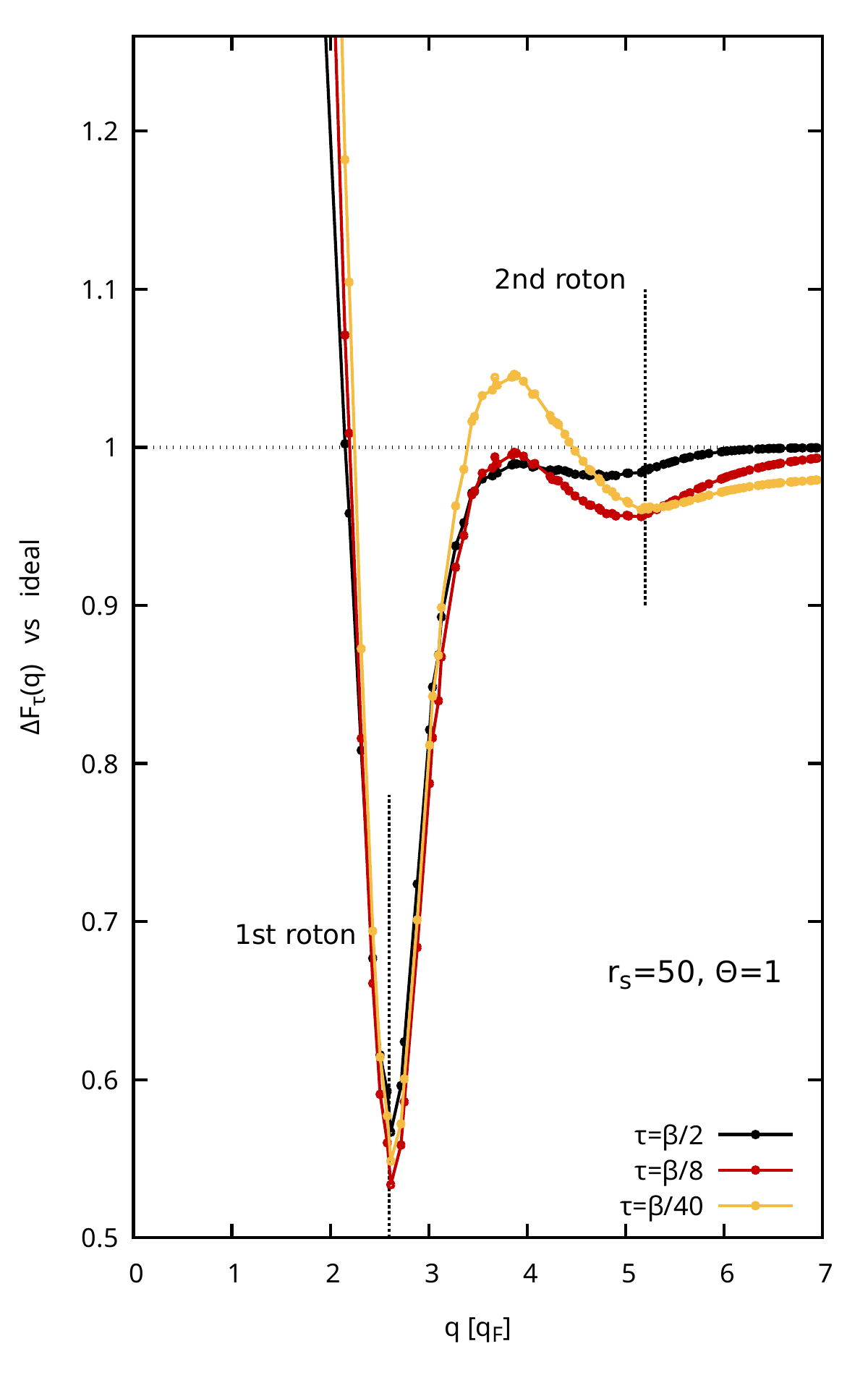}
    \caption{\emph{Ab initio} PIMC results for the ITCF $\tau$-decay measure [Eq.~(\ref{eq:Delta_F_tau})] of the 2DEG for $\tau=\beta/2$ (black), $\tau=\beta/8$ (red) and $\tau=\beta/40$ (yellow), divided by the corresponding $\tau$-decay measure of the ideal 2D Fermi gas at the same conditions. The left and right panels correspond to $r_s=30$ and $r_s=50$, respectively, both at $\Theta=1$.
    }
    \label{fig:1renorm_decay_rs30}
\end{figure*}

\begin{figure}
    \centering
 \includegraphics[width=0.425\textwidth]{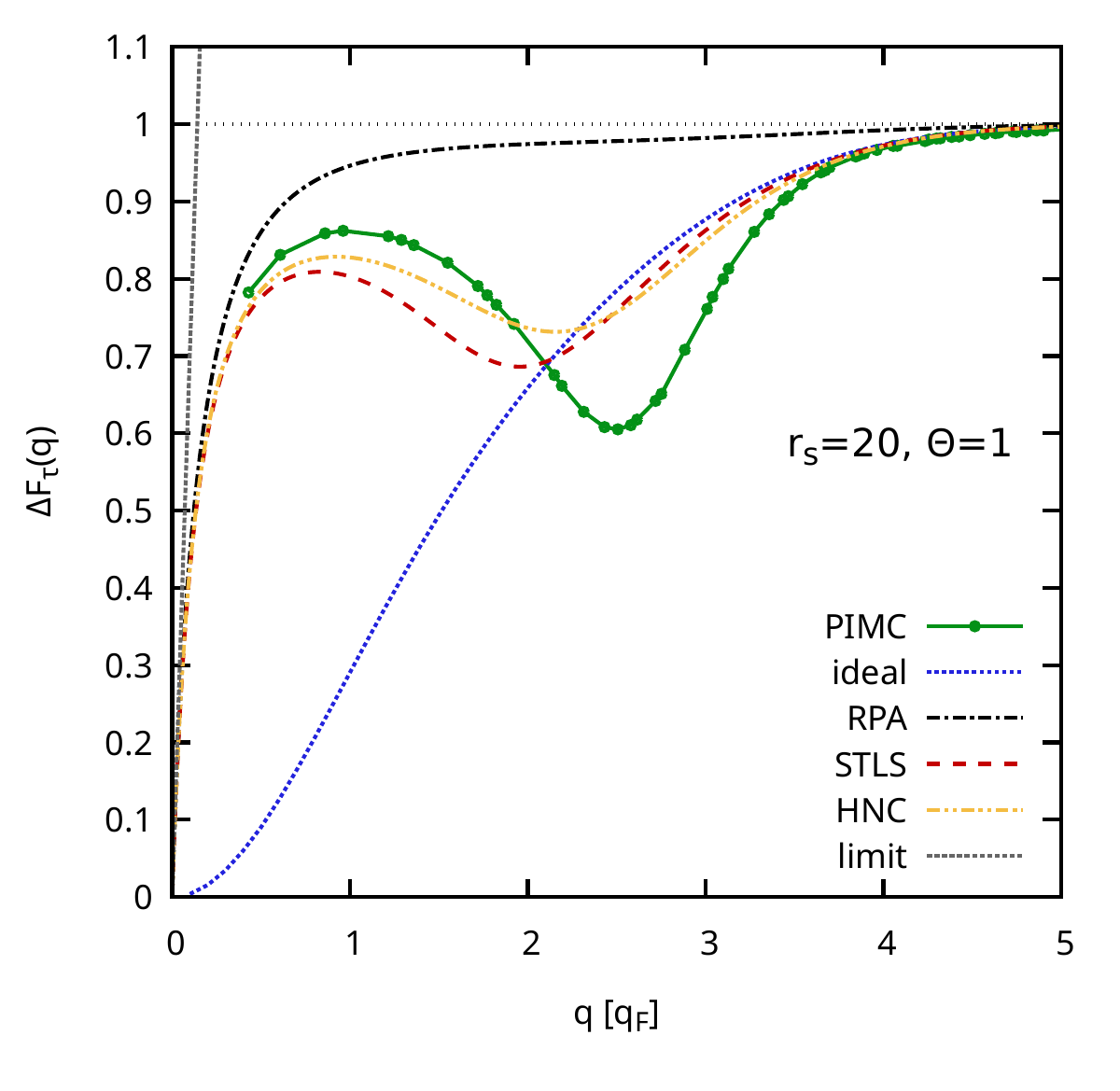}\\\vspace*{-1.2cm} \includegraphics[width=0.425\textwidth]{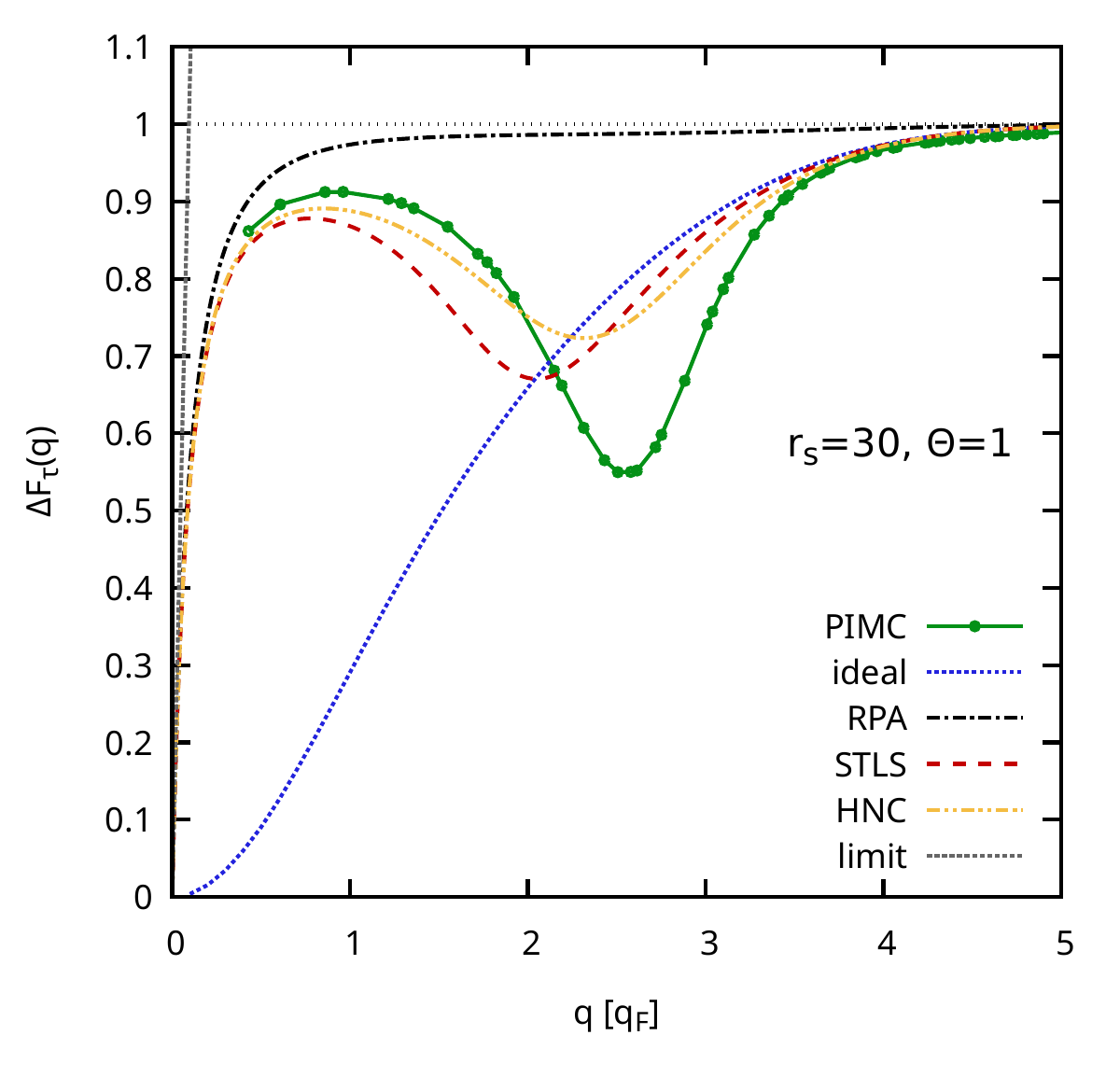}\\\vspace*{-1.2cm} \includegraphics[width=0.425\textwidth]{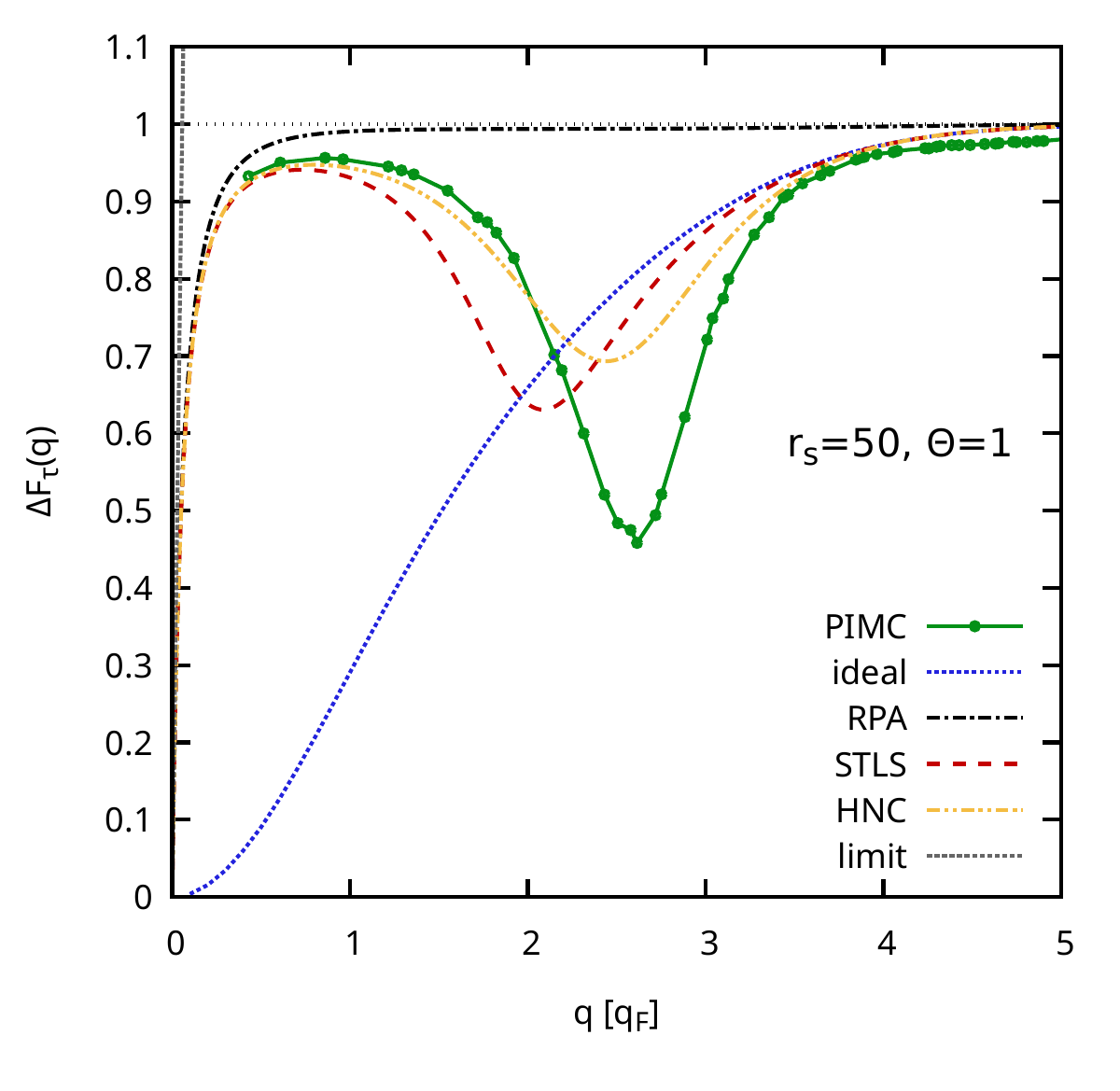}
    \caption{The ITCF $\tau$-decay measure $\Delta{F}_\tau(\mathbf{q})$ [Eq.~(\ref{eq:Delta_F_tau})] of the 2DEG, evaluated for $\tau=\beta/2$ at $\Theta=1$ and $r_s=20$ (top), $r_s=30$ (center), $r_s=50$ (bottom). The different lines distinguish PIMC simulations (solid green) from the ideal 2D Fermi gas (dotted blue), the RPA (dash-dotted black), the STLS scheme (dashed red), and the HNC scheme (dash-double-dotted yellow). The dotted gray curve corresponds to the $q\to0$ limit, Eq.~(\ref{eq:Fotis}).    }
    \label{fig:check_decay_rs30_theta1}
\end{figure}

\subsection{Spectral properties and roton-type feature\label{sec:roton}}

Let us conclude our investigation of the strongly coupled 2DEG with a dedicated analysis of its spectral properties as encoded into the ITCF. Dornheim \emph{et al.}~\cite{Dornheim_MRE_2023} have argued that the main trends of the dynamic structure factor can be grasped directly from $F(\mathbf{q},\tau)$, without the need to numerically invert the bilateral Laplace transform. For example, the shift of spectral weight in $S(\mathbf{q},\omega)$ towards higher frequencies that scales as $\sim q^2/2$ must be reflected by a faster decay in $\tau$, which is exactly what we observe in Figs.~\ref{fig:3D_rs30} and \ref{fig:3D_T}. Conversely, any down shift in spectral weight as it occurs in the vicinity of a roton-type feature in the dispersion relation must manifest in $F(\mathbf{q},\tau)$ as a reduced $\tau$-decay. This is intuitive as strong exchange--correlation effects stabilize correlations along the imaginary-time diffusion process in our PIMC simulations. These considerations are readily quantified by the $\tau$-decay measure,
\begin{eqnarray}\label{eq:Delta_F_tau}
    \Delta F_\tau(\mathbf{q}) = \frac{F(\mathbf{q},0) - F(\mathbf{q},\tau)}{F(\mathbf{q},0)}\ ,
\end{eqnarray}
as shown in Fig.~\ref{fig:decay_rs30} for $\tau=\beta/2$. Specifically, the left panel shows our PIMC results at $r_s=30$ and $\Theta=0.5,\dots,16$. 

For the two highest temperatures, Eq.~(\ref{eq:Delta_F_tau}) is relatively featureless, indicating the absence of pronounced correlation features in the dynamic structure factor. For $\Theta\lesssim4$, we find a pronounced minimum at the expected position of $q\approx2.5q_\textnormal{F}$, which we interpret as a clear signature of the roton consistent with the 3D UEG~\cite{Dornheim_MRE_2023,Chuna_JCP_2025}. The right panel of Fig.~\ref{fig:decay_rs30} shows our PIMC results at $\Theta=4$ and a very broad range of density parameters, $r_s=0.1,\dots,50$. For large $q$, all curves converge towards each other as the dispersion attains the single-particle limit that does not depend on the density. For this value of the reduced temperature $\Theta$, we locate the onset of the roton-type behavior at $q\approx2.5q_\textnormal{F}$ at $r_s\gtrsim20$.

To resolve the roton-type feature more clearly, we follow Chuna \emph{et al.}~\cite{Chuna_JCP_2025}, who suggested to divide $\Delta F_\tau(\mathbf{q})$ by the corresponding results for the ideal Fermi gas. The respective results at $\Theta=1$ are shown in Fig.~\ref{fig:1renorm_decay_rs30} for $\tau=\beta/2$ (black), $\tau=\beta/8$ (red) and $\tau=\beta/40$ (yellow); the left and right panels show results at $r_s=30$ and $r_s=50$, respectively. All curves diverge for $q\to0$ as $\Delta F_\tau(\mathbf{q})$ vanishes more quickly in this limit for the ideal than for the interacting system, cf.~Fig.~\ref{fig:check_decay_rs30_theta1} below. Furthermore, all the curves converge towards unity for large $q$, since the single-particle limit is identical for both the ideal Fermi gas and the 2DEG. We find a clearly reduced relative $\tau$-decay in the interacting compared to the ideal system at $q=2.6q_\textnormal{F}$, which is the "first" roton-type feature. In addition, we find a shallow yet significant second feature at twice the roton wavenumber, which is the incipient "second" roton initially reported in Ref.\cite{Chuna_JCP_2025} for the 3D UEG.
We note that in particular the second roton can be resolved more easily for $\tau\ll\beta/2$, as $\Delta F_\tau(\mathbf{q})$ nearly vanishes for large $\tau$ at these wavenumbers, cf.~Fig.~\ref{fig:FiniteSize1} above.

We conclude the analysis of the 2DEG spectral properties in the imaginary-time domain with a comparison between our quasi-exact PIMC reference results and various dielectric theories in
Fig.~\ref{fig:check_decay_rs30_theta1}. Specifically, we compare PIMC (solid green) to the ideal Fermi gas (dotted blue), the RPA (dash-dotted black), the STLS scheme (dashed red) and the HNC scheme (dash-double-dotted yellow); the top, center and bottom panels show results at $\Theta=1$ and $r_s=20$, $r_s=30$, $r_s=50$, respectively. Overall, we find the same qualitative trends for all three considered cases. The ideal Fermi gas strongly differs from the other curves for large and intermediate wavelengths and only becomes accurate in the single-particle limit of $q\gg q_\textnormal{F}$. The mean-field level RPA constitutes a significant improvement for $q\lesssim q_\textnormal{F}$,
where it attains the exact asymptotic limit of
\begin{eqnarray}\label{eq:Fotis}
    \lim_{q\to0} \Delta F_{\beta/2}(\mathbf{q}) = \frac{\beta^2\pi n}{4}q \ ,
\end{eqnarray}
see the dotted grey curves. Nevertheless, it completely fails to capture the roton-type feature for $q\approx 2.5q_\textnormal{F}$. This is unsurprising given the exchange--correlation origin of this feature and is consistent with previous investigations of the 3D UEG~\cite{dornheim_dynamic,Dornheim_Nature_2022,Chuna_JCP_2025}. The two approximate closures for the local field correction [cf.~Eq.~(\ref{eq:define_LFC}) above] within the STLS and HNC schemes are both capable of providing the correct qualitative behavior, even though substantial and systematic deviations from the PIMC reference results are evident. Interestingly, the STLS gives a somewhat deeper roton feature, whose position is captured more accurately by the HNC. We hope that these findings will help to guide the future development of improved dielectric theories on the static and dynamic level.

\section{Summary and Outlook\label{sec:outlook}}

In this work, we presented our extensive new \emph{ab initio} PIMC results for the two dimensional uniform electron gas, covering a broad range of density parameters ($r_s=0.1,\dots,50$) and temperatures ($\Theta=0.5,\dots,16$). After verifying our implementation by studying the convergence with the number of imaginary-time propagators and comparing against independent configuration PIMC results, we touched upon the fermion sign problem, which constitutes the main limitation of our simulations. Despite the corresponding limit with respect to the feasible system size, we are confident that the wavenumber resolved properties investigated here are hardly affected by the finite simulation cell apart from the discrete $\mathbf{q}$-grid.

From a physics perspective, we studied the static structure factor $S(\mathbf{q})$ and static linear density response $\chi(\mathbf{q})$ across temperature regimes to observe the onset of liquid behavior in the low-density regime upon decreasing the temperature. We found that the pronounced structure in both quantities is not appropriately captured even by the static HNC scheme, which currently constitutes the most sophisticated dielectric scheme for the 2DEG. Furthermore, we presented an in-depth analysis of the spectral properties of the 2DEG from the perspective of the imaginary-time density--density correlation function $F(\mathbf{q},\tau)$, including a brief discussion of the frequency moments of the dynamic structure factor and the f-sum rule. A key finding of our work is the strong evidence for a first roton-type feature at intermediate wavenumbers, which is consistent with previous investigations of the 3D UEG~\cite{dornheim_dynamic,Dornheim_MRE_2023} and 2D helium films~\cite{Godfrin2012}.
Moreover, we found a less pronounced but significant second roton~\cite{Chuna_JCP_2025} at the electronic Fermi temperature and $r_s=30,50$ for twice the wavenumber of the original feature. Finally, we used our quasi-exact PIMC reference data to rigorously assess the quality of dielectric schemes that feature a static local field correction, namely the RPA, STLS and HNC. In contrast to the RPA, both the STLS and HNC qualitatively capture the first roton feature, but substantial deviations remain.

We expect the availability of our highly accurate PIMC results~\cite{repo} to be of value for a plethora of future investigations, including the development of improved static and dynamic dielectric theories~\cite{IIT,tanaka_hnc,Tanaka_CPP_2017,Tolias_PRB_2024,Tolias_JCP_2021,Tolias_JCP_2023,kalkavouras2026dielectricformalism2duniform}, and the benchmarking of novel methods that should be less severely afflicted by the fermion sign problem~\cite{Yilmaz_JCP_2020,Chin_PRE_2015,Hirshberg_JCP_2020,Dornheim_Bogoliubov_2020,Xiong_JCP_2022,Yang_Entropy_2025,Xiong_JCP_2025,Dornheim_JCP_xi_2023,Dornheim_JPCL_2024,Dornheim_PRR_2026}; the latter might also include testing approximate nodal structures for restricted PIMC simulations~\cite{Ceperley1991}. In addition, our ITCF data set can be used as input for the calculation of related properties, including the static and dynamic Matsubara local field correction~\cite{Tolias_JCP_2024,Dornheim_PRB_2024,Dornheim_EPL_2024,MOLDABEKOV2025104144}, various frequency moments of the dynamic structure factor~\cite{Dornheim_moments_2023,Dornheim_MRE_2023,kalkavouras2026kineticenergycubicsum}, and even as input for an analytic continuation to reconstruct the full $S(\mathbf{q},\omega)$~\cite{JARRELL1996133,Chuna_JPA_2025,BENEDIXROBLES2026109904,dornheim_dynamic}. Additional future work might also include PIMC based equation-of-state calculations, which will involve the exploration of appropriate finite-size corrections and of analytical parameterizations.

\begin{acknowledgements}

\noindent This work has received funding from the European Research Council (ERC) under the European Union’s Horizon 2022 research and innovation programme (Grant agreement No. 101076233, "PREXTREME"). Views and opinions expressed are however those of the authors only and do not necessarily reflect those of the European Union or the European Research Council Executive Agency. Neither the European Union nor the granting authority can be held responsible for them. Tobias Dornheim gratefully acknowledges funding from the Deutsche Forschungsgemeinschaft (DFG) via project DO 2670/1-1. Paul Hamann gratefully acknowledges helpful comments by Kai Hunger.

Computations were performed on a Bull Cluster at the Center for Information Services and High-Performance Computing (ZIH) at Technische Universit\"at Dresden and at the Norddeutscher Verbund f\"ur Hoch- und H\"ochstleistungsrechnen (HLRN) under grant mvp00024.
The authors gratefully acknowledge the computing time made available to them on the high-performance computer Otus at the NHR Center Paderborn Center for Parallel Computing (PC2) (ID 27589, HiFi-WDM). This center is jointly supported by the Federal Ministry of Research, Technology and Space and the state governments participating in the National High-Performance Computing (NHR) joint funding program.

\end{acknowledgements}

%%%%%%%%%%%%%%%%%%%%%%%%%%%%%%%%%%%%%%%%%%%%%%%%%%%%%%%%%%%%%%%%%%%%%%%%%%%%%%%%
% literature
%%%%%%%%%%%%%%%%%%%%%%%%%%%%%%%%%%%%%%%%%%%%%%%%%%%%%%%%%%%%%%%%%%%%%%%%%%%%%%%%
\bibliography{bibliography}
\end{document}